\documentclass[%
superscriptaddress,reprint, twocolumns
 amsmath,amssymb,
pre,
]{revtex4-1}

\usepackage{url}
\usepackage{graphicx}% Include figure files
\usepackage{dcolumn}% Align table columns on decimal point
\usepackage{makecell,xfrac}
\usepackage{bm}% bold math
\usepackage{caption,subcaption}
\usepackage{amsmath}
\usepackage{soul}
\usepackage{tocloft}
\usepackage{tabu}
\usepackage{booktabs} \newcommand{\ra}[1]{\renewcommand{\arraystretch}{#1}}
\usepackage{color}

\begin{document}

\title{Analysis and simulation of intervention strategies against bus bunching by means of an empirical agent-based model}% Force line breaks with \\

\author{Wei Liang Quek}
\email{weiliang.quek@ntu.edu.sg}
\affiliation{%
 School of Humanities, Nanyang Technological University, Singapore 639818, Singapore}

\author{Ning Ning Chung}
%\email{nnchung@suss.edu.sg}
\affiliation{Centre for University Core, Singapore University of Social Sciences, Singapore 599494}

\author{Vee-Liem Saw}
%\email{veeliem@ntu.edu.sg}
\affiliation{%
 Division of Physics and Applied Physics, School of Physical and Mathematical Sciences, Nanyang Technological University, Singapore 637371, Singapore}
 \affiliation{Data Science \& Artificial Intelligence Research Centre, Nanyang Technological University, Singapore 639798}

\author{Lock Yue Chew}%
%\homepage{http://www.ntu.edu.sg/home/lockyue/}
\email{lockyue@ntu.edu.sg}
\affiliation{%
 Division of Physics and Applied Physics, School of Physical and Mathematical Sciences, Nanyang Technological University, Singapore 637371, Singapore}
\affiliation{Data Science \& Artificial Intelligence Research Centre, Nanyang Technological University, Singapore 639798}
\affiliation{Complexity Institute, Nanyang Technological University, Singapore 637723}%Lines break automatically or can be forced with \\


\date{\today}% It is always \today, today,

\begin{abstract}
In this paper, we propose an Empirically-based Monte Carlo Bus-network (EMB) model as a test bed to simulate intervention strategies to overcome the inefficiencies of bus bunching. The EMB model is an agent-based model which utilizes the positional and temporal data of the buses obtained from the Global Positioning System (GPS) to constitute: (1) a set of empirical velocity distributions of the buses, and (2) a set of exponential distributions of inter-arrival time of passengers at the bus stops. Monte Carlo sampling is then performed on these two derived probability distributions to yield the stochastic dynamics of both the buses' motion and passengers' arrival. Our EMB model is generic and can be applied to any real-world bus network system. In particular, we have validated the model against the Nanyang Technological University's Shuttle Bus System by demonstrating its accuracy in capturing the bunching dynamics of the shuttle buses. Furthermore, we have analyzed the efficacy of three intervention strategies: holding, no-boarding, and centralized-pulsing, against bus bunching by incorporating the rule-set of these strategies into the model. Under the scenario where the buses have the same velocity, we found that all three strategies improve both the waiting and travelling time of the commuters. However, when the buses have different velocities, only the centralized-pulsing scheme consistently outperform the control scenario where the buses periodically bunch together.
\end{abstract}

%\pacs{PACS numbers: 05.65.+b, 89.40.Bb, 89.75.Fb, 45.70.Vn}

% \keywords{NaSch model, Cellular automaton, Multi-point tollbooth, Lane Expansion, Fundamental diagram}%Use showkeys class option if keyword
                           % display desired
\maketitle
%\tableofcontents
\section{Introduction}

The pervasiveness \cite{Model1964,Gershenson2009,Verbich2016,Saw2019,Chew2019} of bus bunching has been a longstanding problem in public transportation systems. Bus bunching occurs when the distance between two or more buses reduces to near zero. In traffic studies \cite{Model1964, OsunaNewell1972, Newell1974, Abkowitz1986, McLeod1998, Eberlein1999, Zolfaghari2004, Sun2005, Gershenson2009, Daganzo2009, Yu2009, Daganzo2011a, Milla2012, Liu2013, Verbich2016, Andres2017, Wu2017a, Petit2018, Liang2019, Saw2019}, bus bunching is often examined through the headway between two buses, i.e. the forward distance or time of a vehicle to its leading vehicle. Bunching occurs naturally as a consequence of the interaction between buses and passengers \cite{Gershenson2009,Saw2019,Chew2019} after discounting the effects of other traffic. When a leading bus continually spends more time to pick up passengers than a trailing bus, distance between the two buses reduces as the trailing bus travels faster than the leading bus. This causes the buses to bunch. A sustained bunching leads typically to an increased average inter-arrival time between buses and an under-utilised capacity of the trailing bus. Indeed, when two or more buses bunch and move together, they serve essentially as a single unit. This increases the average inter-arrival time between the buses.

In a bid to improve the performance of bus systems, researchers and traffic planners have been proposing intervention methods to alleviate bus bunching. In the 70s, Refs. \cite{Model1964}, \cite{OsunaNewell1972} and \cite{Newell1974} formed the first theoretical framework of an intervention strategy for overcoming bus bunching: a holding strategy. This strategy has been adapted in numerous studies \cite{Abkowitz1986,McLeod1998,Zolfaghari2004,Yu2009}, even very recent ones \cite{Liang2019}. Other solutions such as stop skipping \cite{Sun2005,Liu2013}, expressing \cite{Eberlein1999} and priority signalling \cite{McLeod1998} have also been explored. 
Effectiveness of the proposed interventions are evaluated mainly through macroscopic modelling \cite{Berrebi2018} and simulations as experimentation on a full bus system is extremely costly while a scaled-down experimentation would not capture the actual dynamics of the bus system. Macroscopic modelling is computationally efficient through its mathematical formulation of the macroscopic states of the traffic system. The efficacy of the intervention strategies is then evaluated from their effects on the macroscopic states of the model.

Recently, agent-based modelling (a sub-class of microscopic modelling which applies a bottom-up modelling approach) has been a popular means in the study of various complex systems \cite{Quadrifoglio2008,Moussaid2011,Froese2014,Bobashev2018}. By modelling the microscopic behavior of the individual components of the system, agent-based models capture macroscopic phenomena as they emerge from the interactions of the individual components. This ground-up approach facilitates logical analysis of the relationship between the modelled microscopic behaviors and the resultant macroscopic phenomena, leading to better understanding of the dynamics of the system. 

Most agent-based transportation models simulate only the flow of the vehicles of interest while the rest of the traffic is conveniently modelled as stochastic noise. In the case of a bus system, buses are represented as agents travelling with a cruise velocity. The velocity often follows a mean value measured from the real system and fluctuates within the measured standard deviation. Generality of such models comes at a cost of being less specific. While velocity of a bus does vary stochastically due to the random existence of other vehicles, certain slow down or speed up are however not random. Typically, buses slow down consistently on hairpin turns, upstream of junctions and before traffic lights. Moreover, certain traffic conditions can be specific to a particular bus route. More than often, such specificity plays an important role in influencing the system's dynamics that an intervention cannot be evaluated accurately by omitting the specificity. One can of course mark down and include all such special segments in the modelling of a fixed bus route, however, velocity of a bus depends also on other microscopic details such as driver's behavior and condition of the road. Inclusion of all these microscopic details increases complexity at the cost of generating key insights from the model.

In this paper, we explore the application of Monte Carlo approach to the agent-based transportation modelling. A Monte Carlo process provides an alternative approach to the modelling of velocities of buses on a fixed route. In the model, velocities of buses at different segments of the route are first collected from the real-world system of interest. 
Buses are then simulated as agents travelled at velocities drawn randomly from those empirical distributions of velocities at the corresponding segments of the route. In this way, the collective influence of the system's microscopic details is captured together with the stochasticity of the traffic.

%In this section, the formulation of the Empirically-based Monte Carlo Bus-Network (EMB) model will first be presented, followed by a series of data analytics to derive the required distributions from our case study of the Nanyang Technological University’s (NTU) campus shuttle service. With that, the parameterized model will be validated by comparing the model outputs against empirical trajectories from the NTU's campus shuttle system. Finally, the effects of different intervention strategies will be studies using the fully parameterized EMB model and their respective efficacy will be discussed.

\section{Empirically-based Monte Carlo Bus-Network (EMB) model}
\label{sect:embformulation}

The proposed Monte Carlo model is a data driven model which begins with a data collection process on details of the route and dynamics of the transportation system of interest. Here, we are interested in a closed loop bus service in a university's campus, specifically, Nanyang Technological University's shuttle bus system (NTU-SBS). At a resolution of $\Delta x = 7.5$m, the bus route is mapped into a 1-dimensional space-discretized array of length $L=688$ under periodic boundary conditions (see Fig. 1). $M=12$ of the cells in the array are characterized as `station' cells to model the $12$ unevenly-spaced stations where the campus buses service. Positions of the stations are $x = 62$, $99$, $152$, $205$, $256$, $307$, $353$, $416$, $479$, $548$, $606$ and $667$. We mark these as stops $1$-$12$ respectively.

Buses move along the route and stop to pick up passengers at each bus stop. Depending on the time of the day, there could be one to seven buses serving the route at one time. Global Positioning System (GPS) coordinates of the buses are collected from 23 Aug 2018 to 11 Nov 2018 through the applet published in the web portal: \url{https://baseride.com/maps/public/ntu/}. These raw data were filtered and transformed into 1-D spatio-temporal trajectories i.e. $x_{i,t}$, which are integers within $[1,688]$ for the position of bus $i$ at time $t$. Here, the data are collected at a temporal resolution of $\Delta t = 12$s. Details of the data processing methodology and the spatio-temporal resolution can be found in Appendix \ref{apx:filtering}. Data from weekends and public holidays were excluded as they are not reflective of an average day of shuttle bus operation. 

\begin{figure}[ht]
    \centering
    \includegraphics[width=\columnwidth]{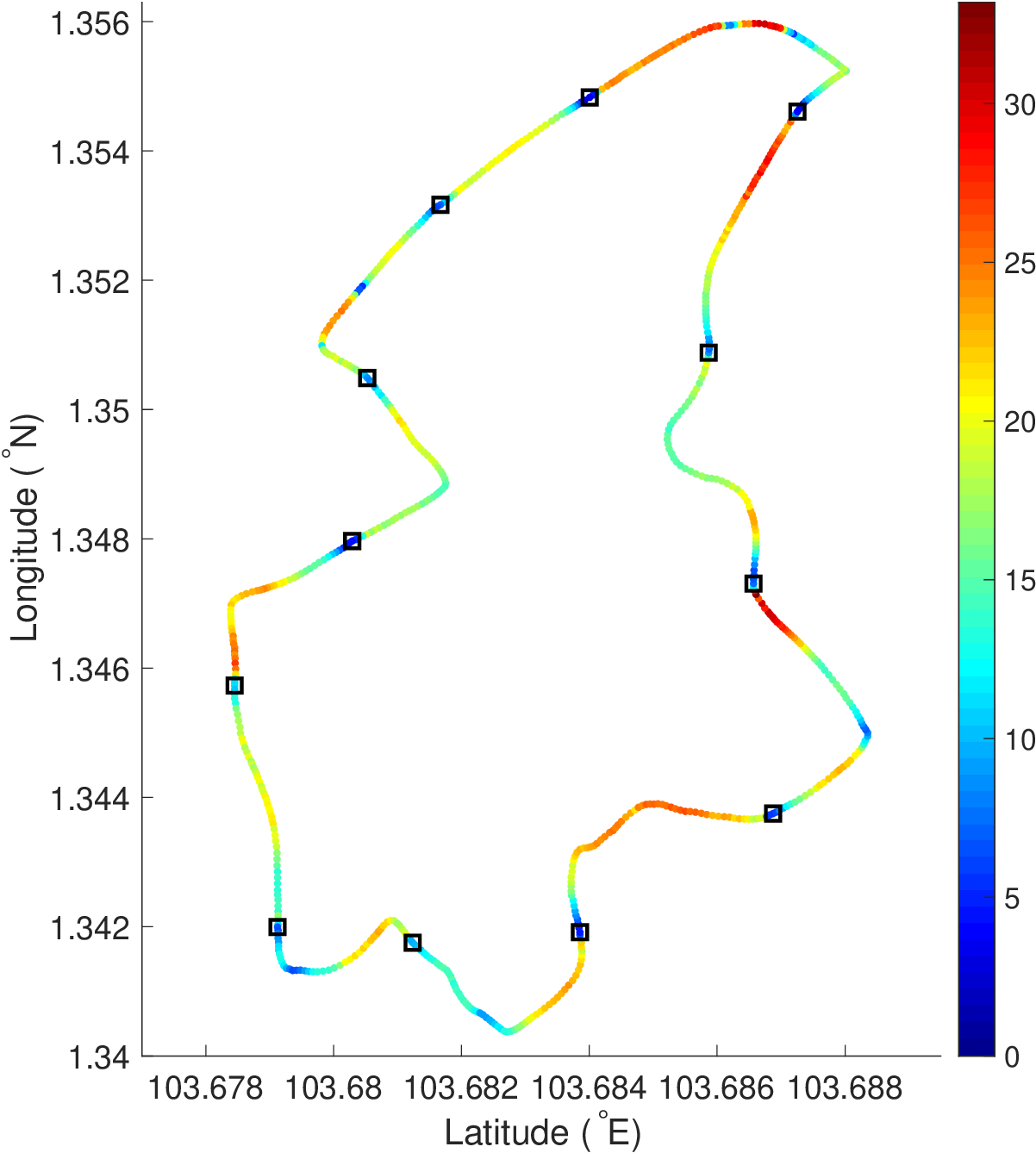}
    \caption{Average velocity of the buses at different positions of the route. The path represents the 688 discretized positions of the route taken by NTU-SBS. The X and Y values represent actual GPS coordinates of the bus route. The color describes the average velocity in units of km/h. Buses move clockwise around the loop.}
    \label{fig:vcolormap}
\end{figure}

\begin{figure}[ht]
     \begin{subfigure}{.7\linewidth}
         \centering
         \includegraphics[width=\textwidth]{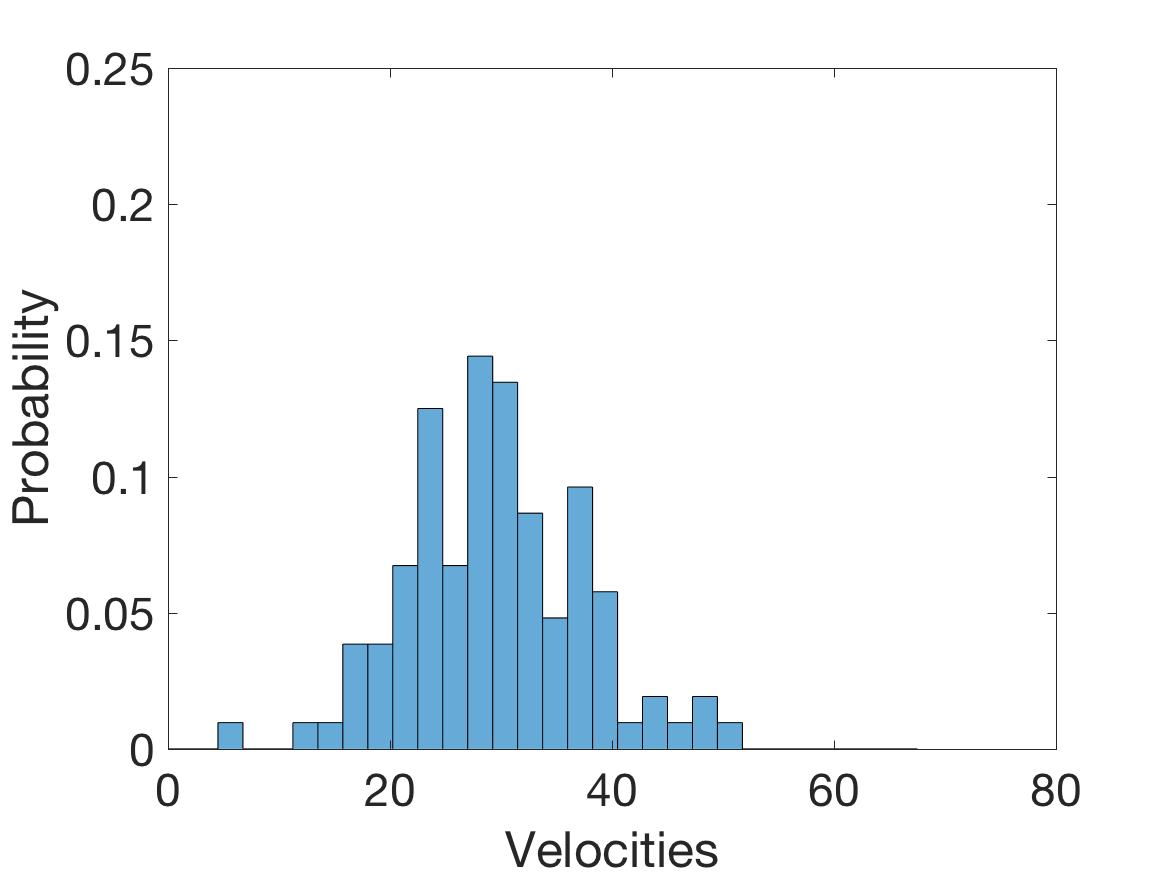}
         \caption{Empirical distribution of velocities of the buses at position $x=542$, which is the middle of a straight stretch of road.}
         \label{fig:vhistfast}
     \end{subfigure}
     \begin{subfigure}{.7\linewidth}
         \centering
         \includegraphics[width=\textwidth]{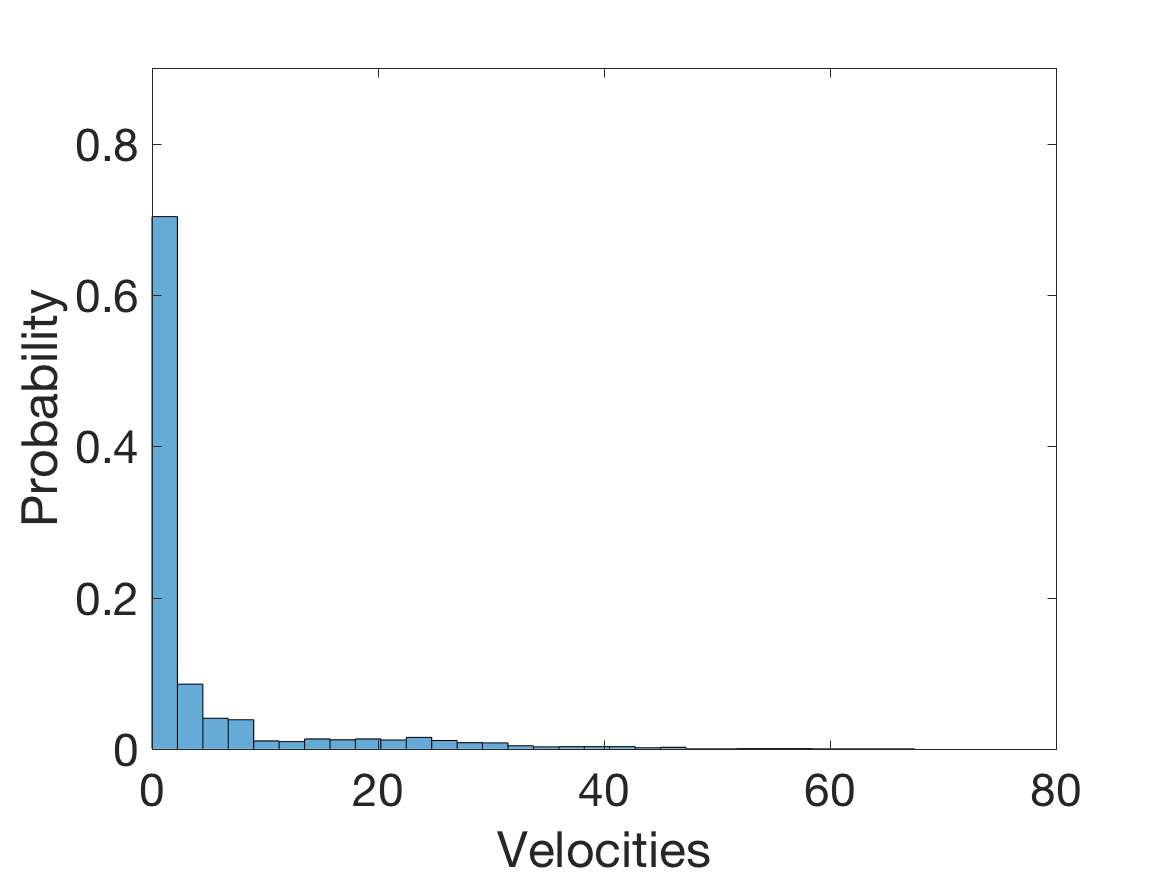}
         \caption{Empirical distribution of velocities of the buses at position $x=61$, which is upstream of a T-junction.}
         \label{fig:vhistslow}
     \end{subfigure}
     \caption{The probability distribution of velocities of the NTU-SBS from 23 Aug 2018 to 11 Nov 2018}
     \label{fig:velocities}
\end{figure}

Velocities of the buses are calculated from the equation: $v_{i,t} = (x_{i,t+1} - x_{i,t})/\Delta t$. For each of the $688$ cells, we collect velocities of buses which pass by the cell. This gives $688$ empirical distributions. The average velocity at which buses cruise on the particular segments of the road can be evaluated from each of the unique distribution at a cell.
Figure \ref{fig:vcolormap} plots the average velocities on different positions along the route. The decrease of average speeds near the bus stops and road junctions are consistent with our heuristic understanding of the system. Figures \ref{fig:vhistfast} and \ref{fig:vhistslow} plot the empirical velocity distribution $P_x(v)$ at the road position where $x = 542$ and $61$, which reflect road segments with the highest and lowest average velocities respectively. Based on our analysis, the average velocity of buses at position $x = 542$ is $31$km/h, which is a reasonable average speed for a bus travelling within a campus. The collected distributions of buses' velocities are to be used in the Monte Carlo process to determine the instantaneous velocities of the buses. Specifically, for each time step $t$, bus $i$ samples a velocity $v^\prime_{i,t}$ from the empirical distribution which corresponds to the bus's position. To account for heterogeneity in the drivers' behaviors, $v^\prime_{i,t}$ is multiplied by a dimensionless constant $K_i$ which scales the average speed of bus $i$. With that, for each time step $t$, bus $i$ moves with an instantaneous velocity of $v_{i,t} = K_i v^\prime_{i,t}$. A systematic method of deriving $K_i$ from empirical data from NTU-SBS will be presented in Sect. 
\ref{sect:validation} of the paper.

In our model, buses can occupy the same position on the route, allowing them to overtake freely. When there are passengers at a bus stop, passing buses will stop at the bus stop to allow passenger boarding. More than one bus is allowed to board passengers at a single bus stop simultaneously and when there are two or more buses, the passenger load is shared equally. When passengers on-board a bus have reached their destinations, the bus will stop to allow for alighting. If there is nobody on the bus who wishes to alight and no passengers at the bus stop waiting to board, the bus will continue travelling towards the next stop. In order to ensure that buses depart from the stop, there is vanishing probability of zero velocity in the empirical velocity distribution for bus departure at the bus stops. Details of the methodology can also be found in Appendix \ref{apx:filtering}.

As the data collected from the web portal does not include passenger boarding or alighting information, these information had to be estimated from the bus trajectories. It is hypothesized that passengers arrive at each bus stop infrequently (ranging from $20$s to $3$mins per passenger). As such, passenger arrival would follow a Poisson process, where the inter-arrival time of passengers would follow an exponential distribution of rate $\lambda$ which describes the rate of passenger arrival. In our Monte Carlo model, passengers are injected to each bus stop $j$ by a Poisson process of mean $\lambda_j$. Each passenger injected at bus stop $j$ will have a randomly assigned destination bus stop $i$ such that $i \neq j$. After injection, each passenger remains at their respective bus stops until the arrival of a bus. When a bus arrives, passengers at a bus stop board on a first-come-first-serve basis. When the approaching bus stop is the destination of any on-board passengers, the bus will stop at the bus stop for passenger alighting. Alighting completes the journey for the passengers. The rate at which boarding and alighting takes place is determined by the parameter $l$ which we set at $1$s. Boarding and alighting are set to take place concurrently, which models how most buses have different doors for boarding and alighting. With this, the stoppage time at a bus stop is to be divided by the loading rate $l$ to yield the number of boarding passengers.

\begin{table}[h!]
\centering
\caption{Mean passenger arrival rate $\lambda_j$ at bus stop $j$ at different times of the day. The unit is given by number of passengers per second.}
\begin{tabular}{|c|c|c|c|c|} 
 \hline
 \thead{Bus stop\\index} & $<$1000h & \thead{1000h - \\1400h} & \thead{1400h - \\2020h} & $>$ 2020h\\ 
 \hline
1 & 0.028 & 0.025 & 0.027 & 0.018 \\
 \hline
2 & 0.022 & 0.019 & 0.015 & 0.009 \\
 \hline
3 & 0.016 & 0.013 & 0.014 & 0.009 \\
 \hline
4 & 0.029 & 0.025 & 0.028 & 0.017 \\
 \hline
5 & 0.016 & 0.012 & 0.010 & 0.006 \\
 \hline
6 & 0.021 & 0.012 & 0.011 & 0.008 \\
 \hline
7 & 0.025 & 0.017 & 0.012 & 0.008 \\
 \hline
8 & 0.046 & 0.027 & 0.018 & 0.009 \\
 \hline
9 & 0.032 & 0.014 & 0.011 & 0.012 \\
 \hline
10 & 0.023 & 0.015 & 0.014 & 0.008 \\
 \hline
11 & 0.028 & 0.015 & 0.012 & 0.006 \\
 \hline
12 & 0.044 & 0.025 & 0.022 & 0.013 \\ 
 \hline
\end{tabular}
\label{table:BSstats}
\end{table}

%While there is no clear definition of Monte Carlo methods \cite{Kalos2008}, the general consensus is that any algorithm bearing the name \emph{Monte Carlo} involves repeated random sampling to obtain a probabilistic interpretation of an expected value. In this study, we follow the definition laid down by \cite{Sawilowsky2003} which defines a Monte Carlo simulation as one which calculates the average behaviour - in this case the average bus trajectory - by drawing a large number of (pseudo) random numbers. 

\label{sect:NTUvel}

%While the EMB model is formulated to model the dynamics of a real bus system, it is unable to model every specific situation that could happen in the real world. Here we highlight two specific situations which will not be considered:(1) The dynamical change(s) brought forth by bus injection or ejection from the system, and (2) the dynamical change(s) brought forth by a sudden change in passenger arrival rate. To circumvent these premises, the simulation methodology are based on specific scenarios where both the number of buses and the passenger arrival rate remain constant.

%For each passenger that has completed their journey, their waiting time is calculated as the time difference between the passenger’s arrival time and boarding time. Similarly, their total trip time is calculated based on the difference between the arrival time and alighting time. 

\section{Model validation}
\label{sect:validation}

In order to validate our model, we split a day into 4 segments with breakpoints at 1000h, 1400h and 2020h. The reason for this split is that bus drivers in NTU shuttle services work in shift which typically start and end at these breakpoints. Such splitting also corresponds to the different passenger arrival rates throughout the day. 

The inter-arrival time of passengers at bus stop $j$, $\Delta t^{\text{arr}}_j$, can be deduced from the time-series of bus arrivals at the bus stop. This requires inputs on the time of arrival of bus $i$ at bus stop $j$, $T^{\text{arr}}_{i,j}$, and the consequential stoppage time $\tau_i$. The departure time of this bus at bus stop $j$ is thus $T^{\text{dep}}_{i,j}=T^{\text{arr}}_{i,j}+\tau_i$. Analogously, the departure time of the previous bus, bus $i-1$, at bus stop $j$ is $T^{\text{dep}}_{i-1,j}=T^{\text{arr}}_{i-1,j}+\tau_{i-1}$. Assuming a uniform passenger arrival rate, we expect
\begin{equation}
\Delta t^{\text{arr}}_j
= \frac{T^{\text{dep}}_{i,j}-T^{\text{dep}}_{i-1,j}}{\tau_i l} \,,
\label{eqn:t_inj}
\end{equation}
where $l$ is the loading rate and $\tau_i l$ being the total number of passengers boarding bus $i$. From the ensemble of bus trajectories, we then form the empirical distribution of $\Delta t^{\text{arr}}_j$ for each bus stop $j$ at each of the four time periods. We observe that the empirical distribution takes the form of an exponential distribution. The mean passenger arrival rate $\lambda_j$ is determined from fitting the respective empirical distribution to the exponential distribution. The results are tabulated in Table \ref{table:BSstats}.

Note that this methodology assumes that all bus stoppages are solely due to passenger movement and that buses moves off immediately after all passengers have completed boarding and alighting. In reality, this is not true. As buses typically stop at bus bays, buses would require a clear traffic on the road before they are able to continue their journey. Also, we have observed that passengers who alight from buses typically cross the road in front of the stationary buses, further increasing its delay in moving off. As such, estimating the number of boarding passengers based on stoppage time results in an over-estimate in the derived $\lambda_j$.

The derived passenger arrival rates were analyzed against heuristic understanding of NTU-SBS. Temporally, the decrease in passenger arrival rate after 2020h is consistent with empirical observations as the majority of students and staff do not reside on campus. Spatially, bus stop 4 seems to consistently have higher arrival rates than other bus stops. This bus stop is located at the \emph{North Spine Plaza}, where there is an aggregation of a variety of amenities. Notably, this bus stop is also located in the vicinity of the biggest library in NTU, and there is usually a significant number of library users up till its closing time at 2200h. This could lead to a relatively high arrival rate, even after 2020h. The highest recorded $\lambda_j$ occurs at bus stop 12 in the morning. This bus stop serves one of the largest residential halls on campus, which could explain the large number of passenger arrival rates in the morning. 

From 13 weeks of segmented data, we select 256 sets of trajectories which start and end when either one of the buses in service ends its shift or a new bus joins the service. These trajectories last from 30 to 84 minutes, involving a fixed number of buses which ranges from 2 to 7. Note that none of the 256 sets of trajectories cross any of the breakpoints. With these, we are able to examine the validity of the EMB model under the conditions when the number of buses are constant, and the mean passenger arrival rates can be described by a set of $\lambda_j$. As an example, one of the trajectories happen on 11 Sept, 14:04h - 15:16h with three buses of average velocities 19.5, 19.7 and 21.3 (in units of km/h). These average velocities $V_i$ are used to estimate the corresponding $K_i$ via $K_i = V_i/V$, where $V$ is the average velocity and it is determined from the empirical distributions across the entire route. The $\lambda_j$ to be used to simulate the trajectory is based on the start-time of the scenario (in this example, 14:04h). This start-time is matched against Table \ref{table:BSstats} to determine the corresponding passenger arrival rate $\lambda_j$ to be employed. 

%the same analysis should be done separately for peak and off-peak hours of a day or even for every hour of the day.

%By utilizing each set of trajectories as a simulation scenario for the EMB model, the passenger arrival rate can be fixed throughout the simulation.

We simulate the 256 trajectories using the EMB model. Each simulation begins with $N$ buses at their respective initial positions, with no passengers on-board. There are also no passengers at any of the $M$ bus stops at this point. Preliminary analysis done by varying the number of initial passengers found similar simulation outcomes even when we start with a reasonable number of passengers on-board or at the bus stop. We then examine the phenomenon of bus bunching by analyzing the relative positions among the $N$ trajectories. Specifically, we adapt the idea of oscillators synchronisation to characterize the phenomenon of complete bus bunching by means of the degree of synchronisation \cite{Saw2019,English2008}:
\begin{equation}
    r^2(t) = \frac{1}{N^2}\bigg[\Big(\sum_i{\cos{\theta_i(t)}} \Big)^2 + \Big(\sum_i{\sin{\theta_i(t)}} \Big)^2 \bigg] \,,
    \label{eqn:r^2}
\end{equation}
where $\theta_i(t)$ is the angular position of oscillator $i$. When applied to our bus system, $\theta_i(t)$ is the angular position of bus $i$. This quantity $r^2$ has values from $0$ to $1$, inclusive. The maximum $r^2=1$ implies complete bunching, where all the buses are bunched as a single platoon. On the other hand, $r^2=0$ describes a state where all the buses are randomly spread out across the whole bus route. Notably, $r^2=0$ does not necessarily describe an absence of bus bunching. Consider for example a case in which four buses bunch up in pairs. If these two pairs occupy antipodal positions, we have $r^2=0$. In that regard, $r^2=1$ serves as a condition for {\it complete} bus bunching, whilst $r^2=0$ can imply no bunching at all; or subsets of bunched buses where these subsets are spread out. Nonetheless, qualitative comparison between simulated passengers' total travelling time and the $r^2$ of the system (presented in Sect. \ref{sect:intervention}) shows that $r^2$ is indicative of the performance of a bus system.

\begin{figure*}[ht]
    \centering
    \begin{subfigure}{.8\linewidth}
        \includegraphics[width=\textwidth]{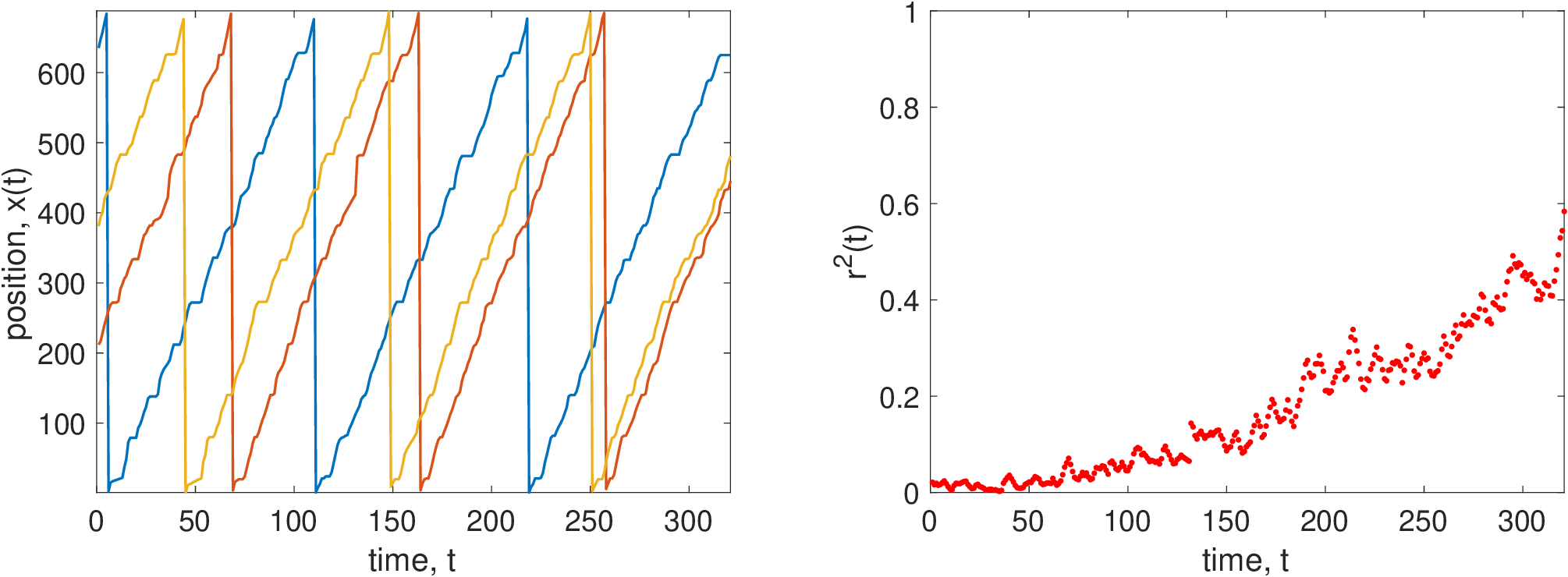}
        \caption{Empirical}
    \end{subfigure}
    \begin{subfigure}{.8\linewidth}
        \includegraphics[width=\textwidth]{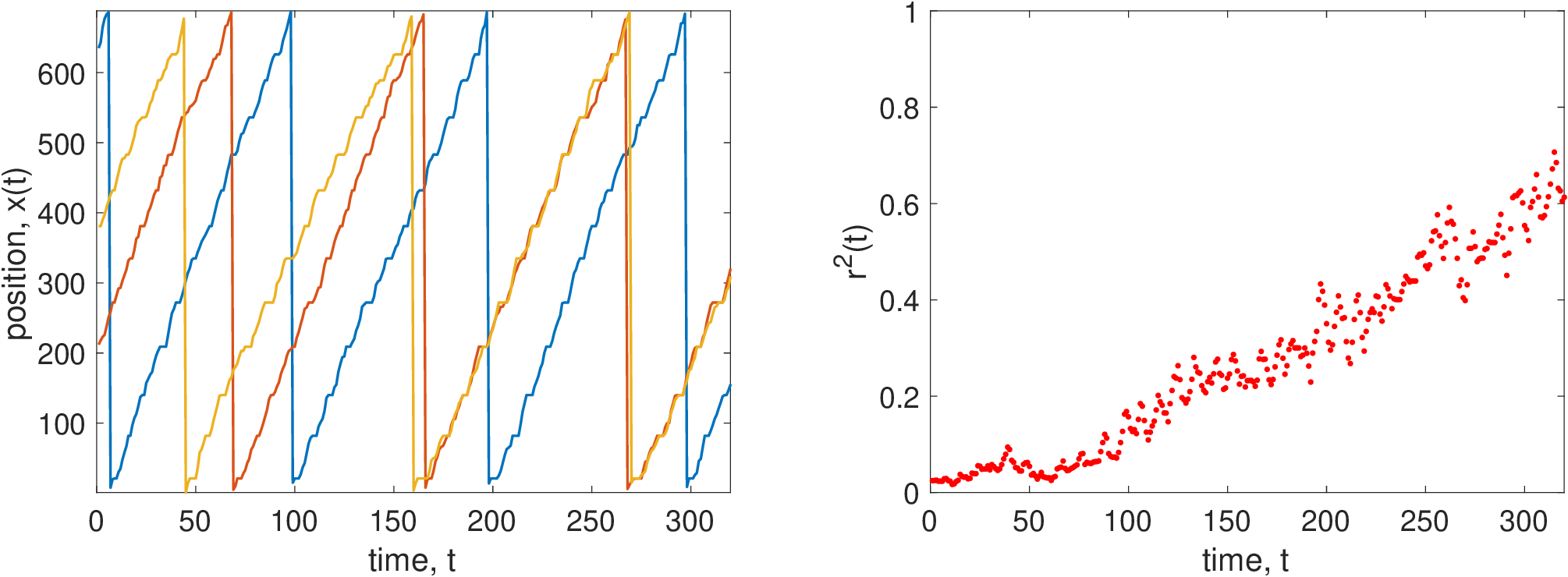}
        \caption{Simulated}
    \end{subfigure}
    \caption{Empirical trajectories of the three buses operating on 11 Sept 18, from 1404h - 1516h. The $r^2$ time series is derived from the trajectories and Eq. \ref{eqn:r^2}.}
    \label{fig:emptraj}
\end{figure*}

In Fig. \ref{fig:emptraj}, we plot the simulated trajectories from a single realization of the EMB model for comparison against the corresponding empirical trajectories of 11 Sept, 14:04h - 15:16h. As expected of a Monte Carlo model, the simulated trajectories do not match the empirical trajectories exactly. However, the simulated trajectories do capture a similar trend of complete bunching as the empirical trajectories, which is clearly illustrated by the simulated $r^2$ ($r_\text{emp}^2$) and empirical $r^2$ ($r_\text{emp}^2$) respectively.

\begin{figure*}[ht]
     \centering
     \begin{subfigure}{.3\linewidth}
         \centering
         \includegraphics[width=\textwidth]{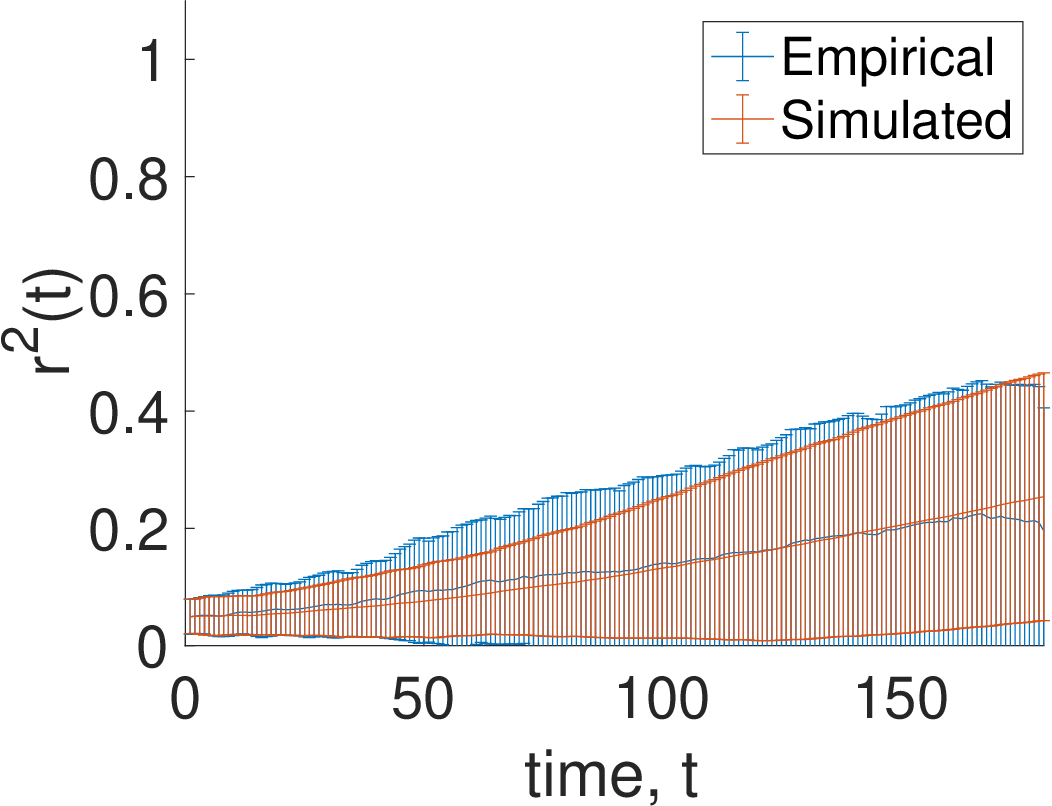}
         \caption{Initial $r^2 \in \big[0.0,0.1\big)$}
         \label{fig:all2all1}
     \end{subfigure}
     \begin{subfigure}{.3\linewidth}
         \centering
         \includegraphics[width=\textwidth]{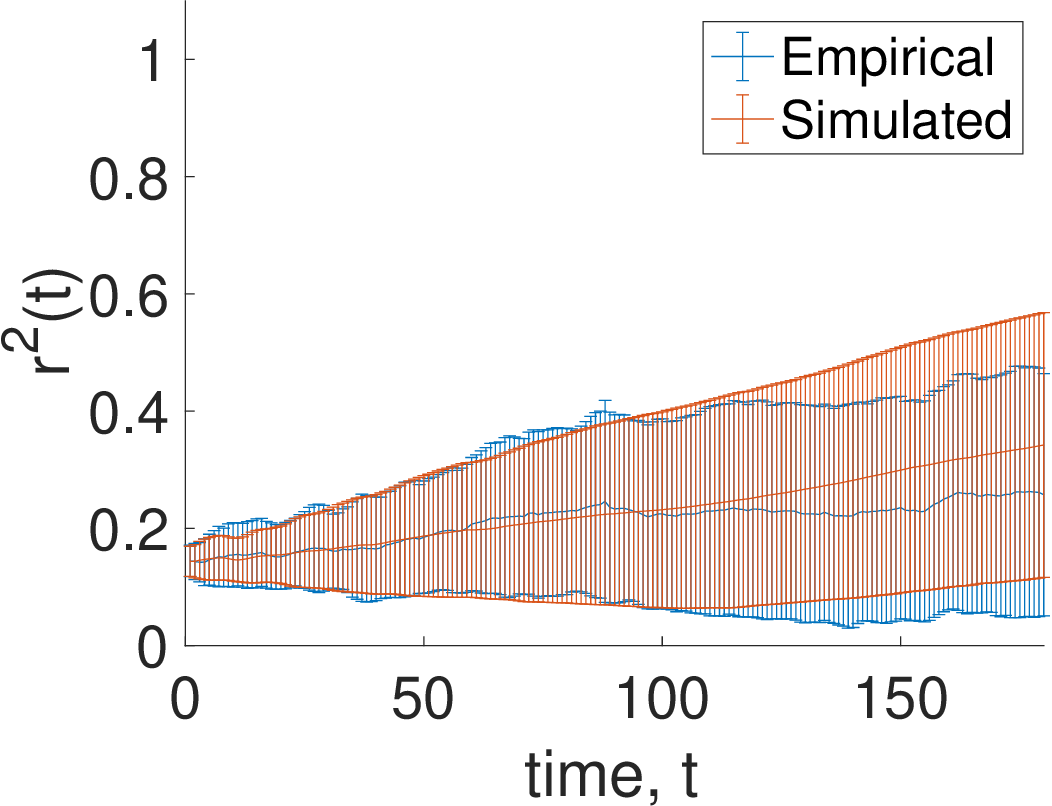}
         \caption{Initial $r^2 \in \big[0.1,0.2\big)$}
     \end{subfigure}
     \begin{subfigure}{.3\linewidth}
         \centering
         \includegraphics[width=\textwidth]{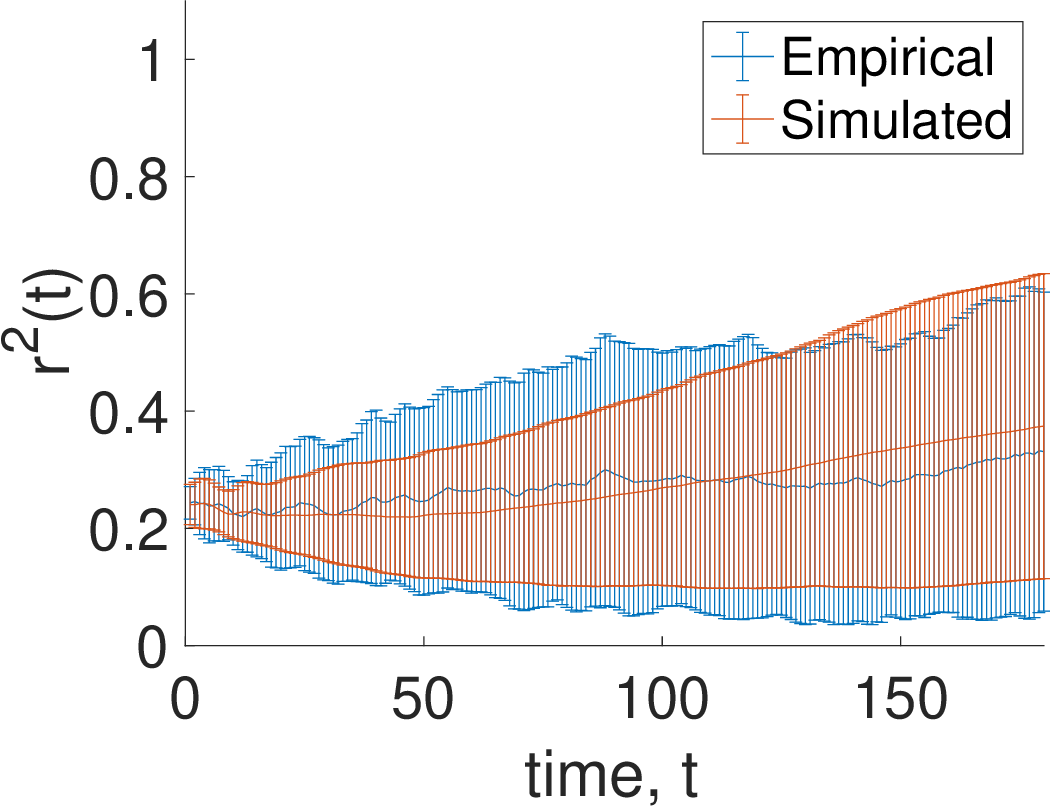}
         \caption{Initial $r^2 \in \big[0.2,0.3\big)$}
     \end{subfigure}
     \begin{subfigure}{.3\linewidth}
         \centering
         \includegraphics[width=\textwidth]{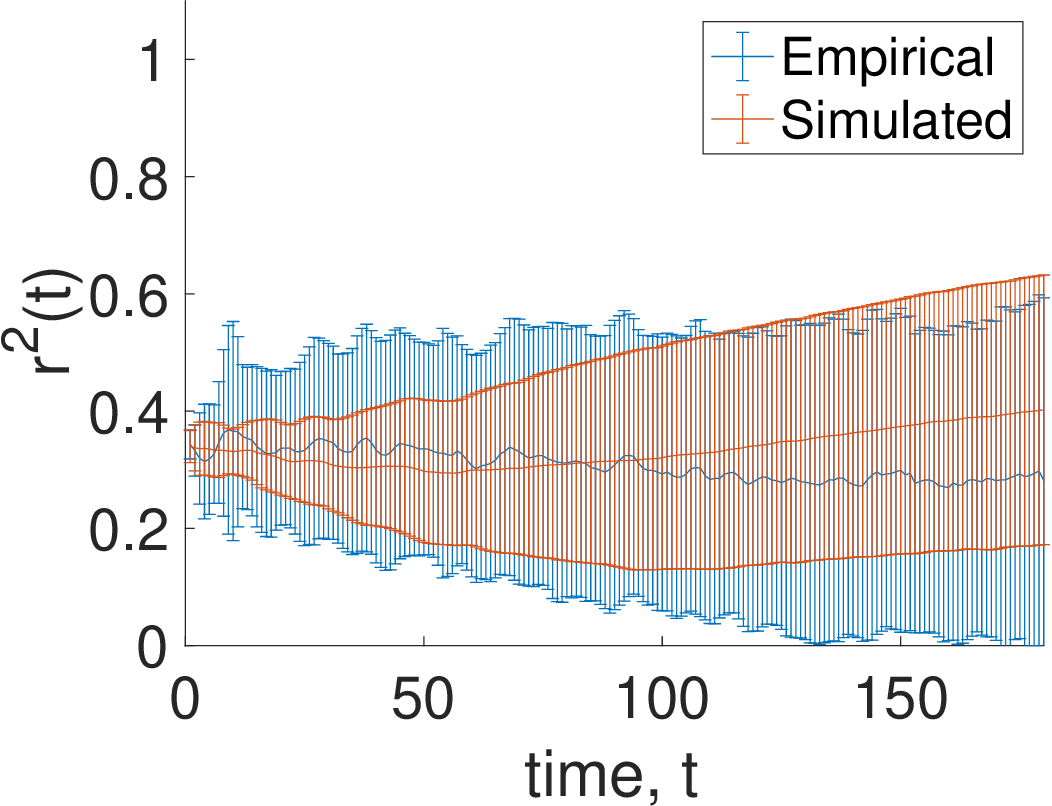}
         \caption{Initial $r^2 \in \big[0.3,0.4\big)$}
     \end{subfigure}
     \begin{subfigure}{.3\linewidth}
         \centering
         \includegraphics[width=\textwidth]{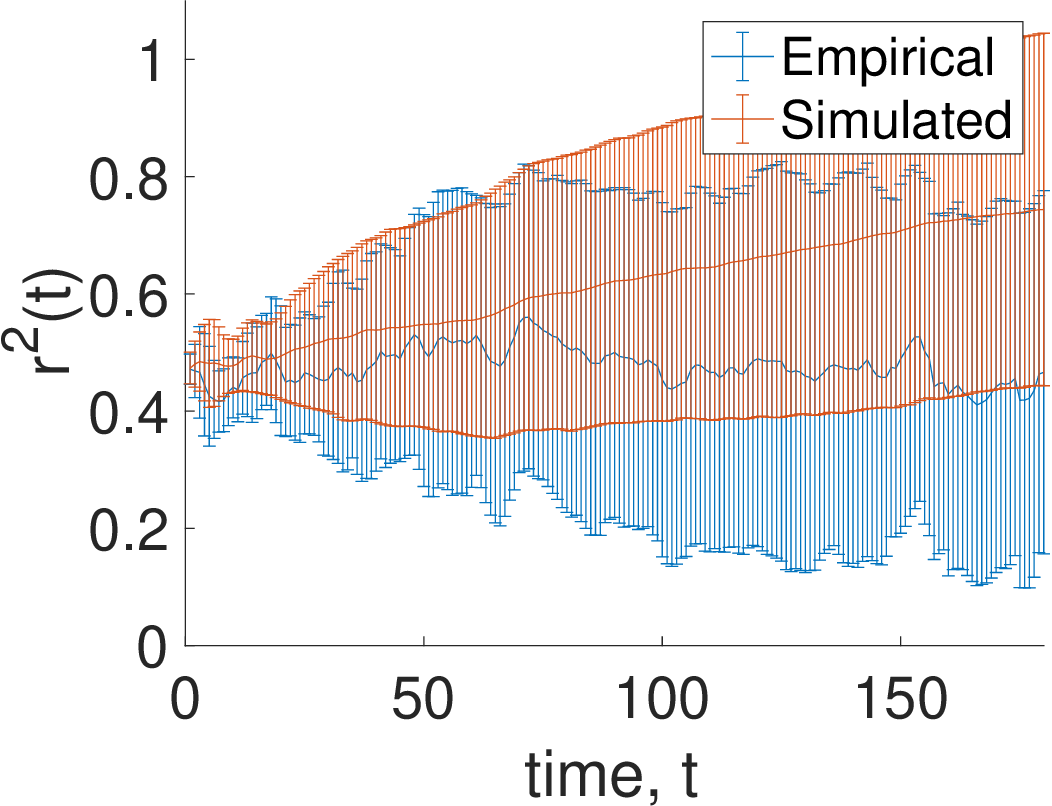}
         \caption{Initial $r^2 \in \big[0.4,0.5\big)$}
     \end{subfigure}
     \begin{subfigure}{.3\linewidth}
         \centering
         \includegraphics[width=\textwidth]{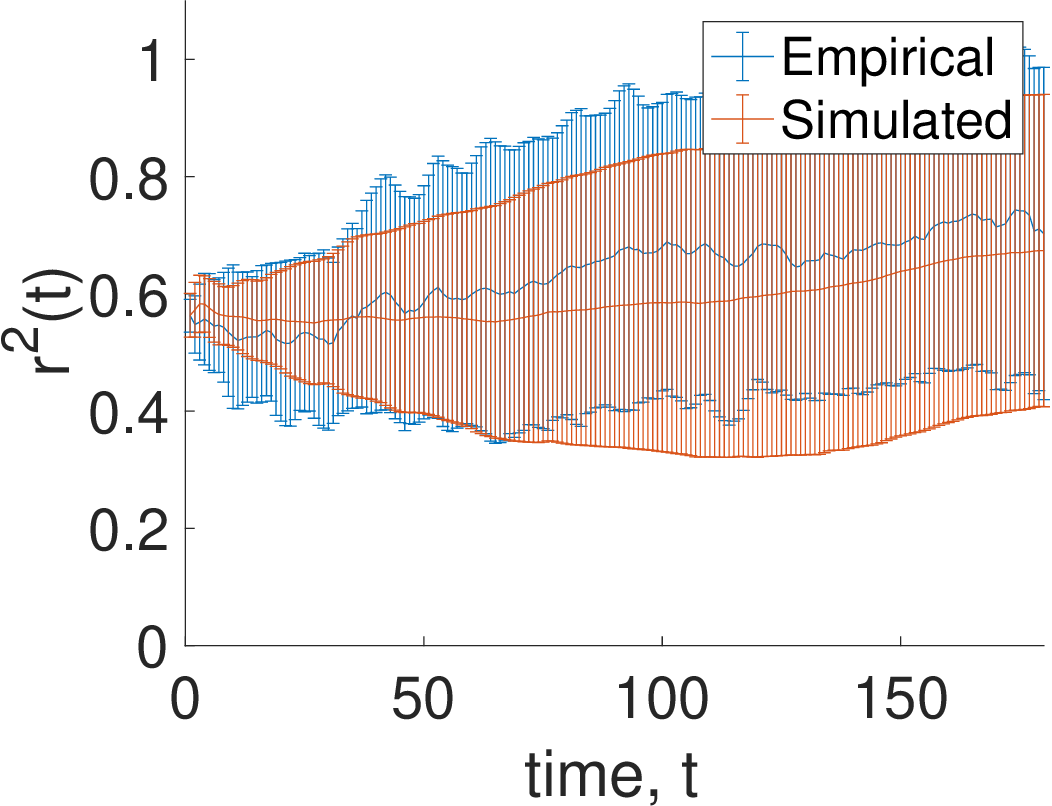}
         \caption{Initial $r^2 \in \big[0.5,0.6\big)$}
     \end{subfigure}
     \begin{subfigure}{.3\linewidth}
         \centering
         \includegraphics[width=\textwidth]{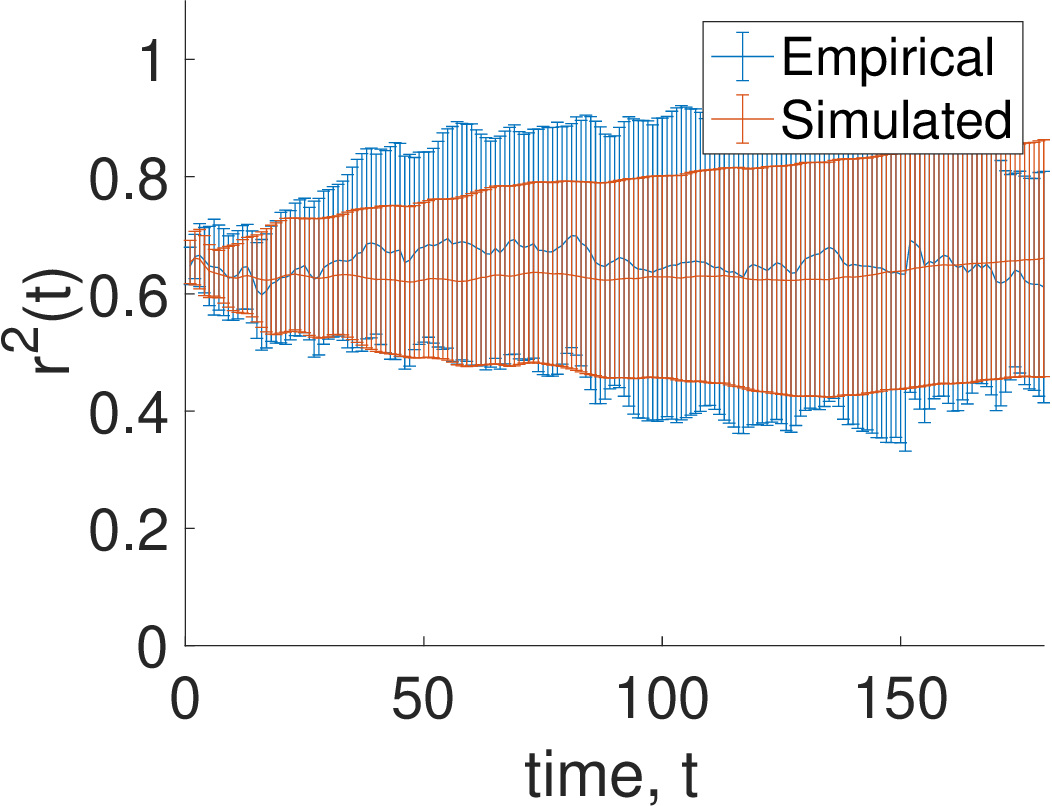}
         \caption{Initial $r^2 \in \big[0.6,0.7\big)$}
     \end{subfigure}
     \begin{subfigure}{.3\linewidth}
         \centering
         \includegraphics[width=\textwidth]{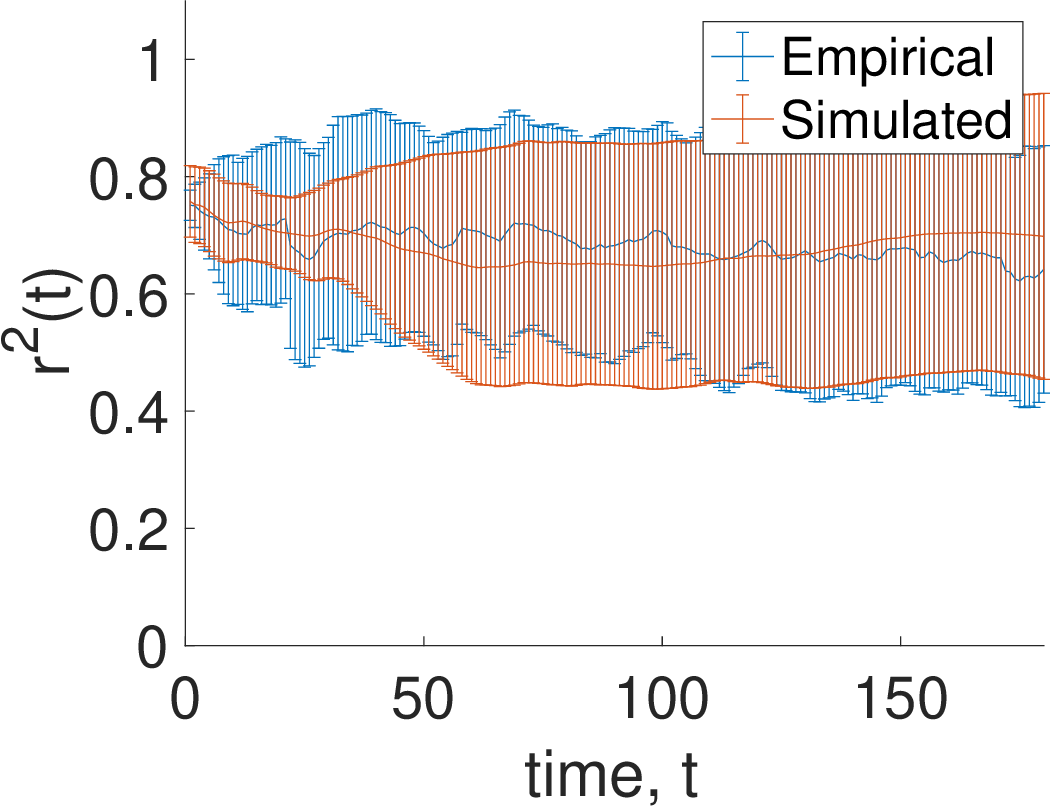}
         \caption{Initial $r^2 \in \big[0.7,0.8\big)$}
     \end{subfigure}
     \begin{subfigure}{.3\linewidth}
         \centering
         \includegraphics[width=\textwidth]{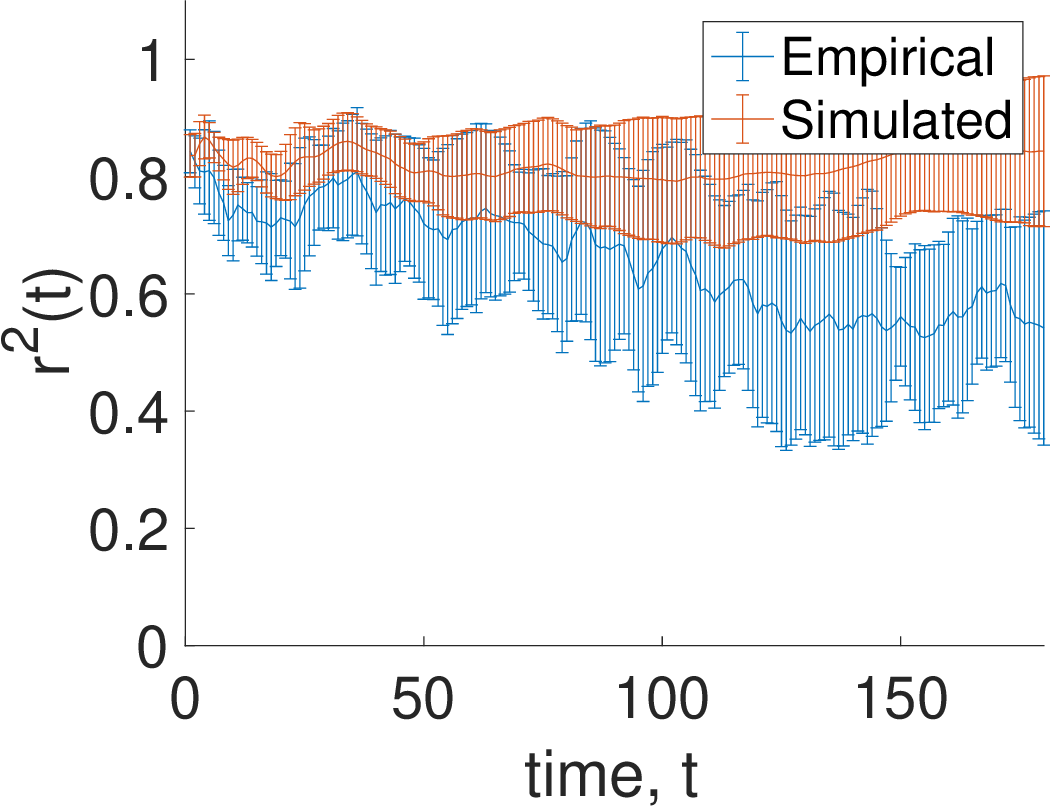}
         \caption{Initial $r^2 \in \big[0.8,0.9\big)$}
         \end{subfigure}
     \begin{subfigure}{.3\linewidth}
         \centering
         \includegraphics[width=\textwidth]{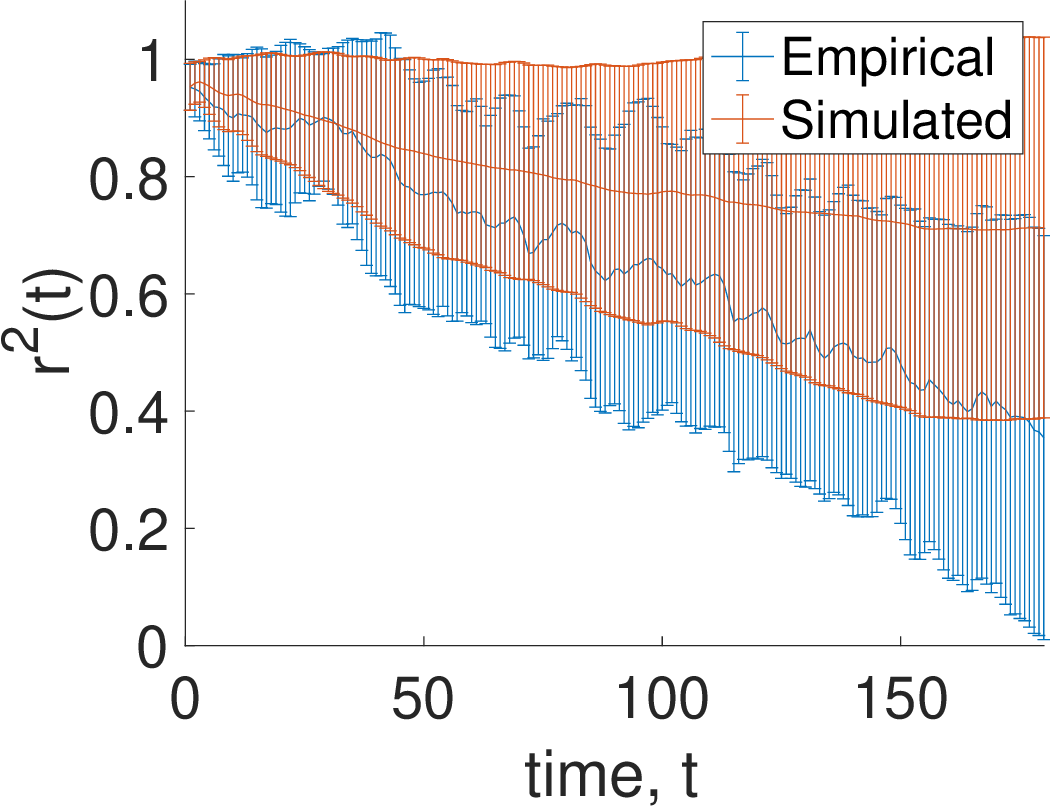}
         \caption{Initial $r^2 \in \big[0.9,1.0\big]$}
         \label{fig:all2all10}
     \end{subfigure}
     \caption{Plot of empirical and simulated trajectories. Each graph is selectively aggregated based on the initial $r^2$ of the trajectories from \big[0.0, 0.1\big) to \big[0.9, 1.0\big]. The statistical similarity between the empirical and simulated trajectories is tabulated in Table \ref{table:all2all}}
     \label{fig:all2all}
\end{figure*}

In our comparison between the model and empirical data, we selectively aggregate the empirical $r^2(t)$ based on their initial $r^2$, i.e. $r_\text{emp}^2(0)$. In other words, we sort the 256 trajectories into 10 different bins based on their initial $r^2$, i.e. $r_\text{emp}^2(0) \in \big[ 0.0, 0.1 \big), \ldots, r_\text{emp}^2(0) \in \big[ 0.9, 1.0 \big] $. Doing so reduces the statistical fluctuation in the empirical trajectories and highlights the general trend on how different states of bunching evolve over time from their initial values. The aggregated empirical trajectories (denoted as $\overline{r^2_\text{emp}}$) (Fig. \ref{fig:all2all}, in blue) shows a general trend of $r^2$ increasing over time, especially for low initial $r^2$. This demonstrates the propensity of the buses to bunch over time. This trend breaks down, however, for the cases of large initial $r^2$. On average, $r_\text{emp}^2$ trajectories which start with buses in an almost completely bunched state tend to unbunch over time.

A similar aggregation is performed on the simulated trajectories, with Fig. \ref{fig:all2all} comparing $\overline{r^2_\text{emp}}$ and $\overline{r^2_\text{sim}}$ trajectories. The statistical similarity of these trajectories is quantified by comparing the average Euclidean distance of the aggregated trajectories (defined by $\frac{\sum^T_{t=1} |\overline{r^2_\text{sim}}(t) - \overline{r^2_\text{emp}}(t) |}{T}$) with the average standard deviation of the aggregated trajectories (Table \ref{table:all2all}) \cite{Chandra2013a,Andres2017}. With exception of the $r_\text{emp}^2(0) \in \big[ 0.8, 0.9 \big)$ bin, the distance of all aggregated trajectories was found to be within the empirical horizon. This reflects a close correspondence between the dynamical behaviour found in the empirical system and that captured by the EMB model.

\begin{table}[h!]
\centering
\caption{Average Euclidean distance between the aggregated simulated and empirical trajectories, tabulated with the standard deviation of both trajectories.}
\begin{tabular}{||c|c c c||} 
 \hline
\thead{Range of \\ initial $r^2$} & \thead{Average\\Euclidian\\distance} & \thead{Average SD\\(data)} & \thead{Average SD\\(simulation)} \\ [0.5ex] 
 \hline\hline
0.0 - 0.1 & 0.01 & 0.14 & 0.11\\ 
0.1 - 0.2 & 0.03 & 0.14 & 0.14\\ 
0.2 - 0.3 & 0.03 & 0.19 & 0.16\\ 
0.3 - 0.4 & 0.05 & 0.22 & 0.16\\ 
0.4 - 0.5 & 0.14 & 0.25 & 0.22\\ 
0.5 - 0.6 & 0.05 & 0.22 & 0.21\\
0.6 - 0.7 & 0.03 & 0.20 & 0.16\\ 
0.7 - 0.8 & 0.03 & 0.19 & 0.18\\ 
0.8 - 0.9 & 0.16 & 0.14 & 0.09\\ 
0.9 - 1.0 & 0.14 & 0.22 & 0.21\\ [1ex] 
 \hline
\end{tabular}
\label{table:all2all}
\end{table}

\section{EMB model as a test bed for intervention strategies in NTU Shuttle Bus System}
\label{sect:intervention}

In this section, the EMB model is used as a test bed for three classes of intervention strategies, with the aim of gaining insights into their effectiveness on alleviating bus bunching. Simulations are initialized with the buses at the equal-headway state and all simulations are only measured at steady state, where the statistics of the system remains stationary in the long time limit. The measurements include the $r^2(t)$ of the buses, the average waiting time and the average travelling time (defined as the sum of the waiting time and the total time spent in the bus) of the passengers. These measurements are compared against a control, which is a similar system that does not involve any interventions. To minimize the effects of statistical variation, the same set of passenger arrival rates are used across all the simulations. 

Depending on the passenger arrival rates $\lambda_j$ and the velocities of the buses, we consider four different scenarios:
\begin{enumerate}
    \item Lull-same -- Low passenger arrival rate and buses have the same intrinsic velocity.
    \item Lull-different -- Low passenger arrival rate and buses have different intrinsic velocities.
    \item Busy-same -- High passenger arrival rate and buses have the same intrinsic velocity.
    \item Busy-different -- High passenger arrival rate and buses have different intrinsic velocities.
\end{enumerate}

From the 256 empirical dataset, we identify 3 scenarios which match the descriptions above:
\begin{itemize}
    \item Lull-same -- 01 Oct 2018, from 21:41h to 22:33h; 2 buses with the same velocity of 15.6 km/h; average arrival rates across the bus stops is $0.011$.
    \item Lull-different -- 10 Oct 2018, from 20:22h to 20:59h; 2 buses with velocities 15.0 km/h and 19.5 km/h respectively; average arrival rates across the bus stops is $0.011$.
    \item Busy-different -- 17 Sept 2018, from 08:56h to 09:32h; 7 buses with velocities 13.3 km/h, 15.9 km/h, 18.0 km/h, 16.9 km/h, 17.1 km/h, 16.0 km/h, 20.4 km/h respectively; average arrival rates across the bus stops is $0.040$.
\end{itemize}
Note that as there is no empirical scenario of a busy phase with buses travelling with the same intrinsic velocity, the busy-same scenario is a hypothetical variant of the busy-different scenario which uses all the parameters of busy-different scenario, but with all the buses having the same intrinsic velocity.

\subsection{Strategy formulation and its effects on the same-velocity scenarios}
\label{sect:intsame}
\subsubsection*{Intervention: Stop-based Holding}
A simple holding strategy involves holding buses at a specified control point for a set amount of time, based on its current time-headway \cite{Berrebi2018}. As a trailing bus tends towards bunching with its leading bus, this strategy holds the trailing bus such that it extends its headway from the leading bus. This produces a negative feedback which opposes the positive feedback mechanism of bus bunching. In a naive headway holding strategy, the holding time of bus $i$ at time $t$ is given by:
\begin{align}
    t^{h\prime}_{i,t}= 
\begin{cases}
    h_t^\ast - h_t, & \text{if } h_t<h_t^\ast\\
    0, & \text{otherwise}
\end{cases} \,,
\end{align}
where $h_t$ is the current time-headway of a bus, and $h_t^\ast$ is a predetermined scheduled headway. For the purposes of minimizing bus bunching in our closed-loop system, the equal time-headway state is chosen as the scheduled headway in our study. This is given by $h_t^\ast = H_t/N$, where $H_t$ is the average period of a bus around the loop, and $N$ is the number of buses. We note that scheduling buses for holding strategy is an ongoing research area \cite{Quadrifoglio2009,Chandra2013a} and using a fixed time interval of $H_t/N$ is a simplification. 

In most holding strategies, control points are set at some - if not all - of the bus stops \cite{Wu2017a,Liu2018,Berrebi2018,Gershenson2009}. In these strategies, $h_t$ is calculated by the time difference between the current time ($t^\text{curr}$) of the bus and the latest departure time at bus stop $j$, i.e. $\text{max}(t^{\text{dep},j})$, before $t^\text{curr}$. Mathematically, it is $h_t = t^\text{curr} - \text{max}(t^{\text{dep},j})$. By applying this measurement in the naive holding strategy, buses only leave the bus stops when its headway is equal or greater than $H_t/N$, i.e. $h_t \geq h_t^\ast = H_t/N$. As this manner of measuring headway is based on when a bus leaves the bus stops, Ref. \cite{Andres2017} refers to it as stop-based headway. In that regard, we refer to this holding strategy as stop-based holding strategy or s-holding.

Reference \cite{Daganzo2009} modified the naive headway approach by multiplying the holding time by a constant $\alpha$ to compensate for unstable headway dynamics found in the real world. With that, holding time of bus $i$ at time $t$ would be:
\begin{align}
    t^h_{i,t}= 
\begin{cases}
    \alpha (h_t^\ast - h_t),& \text{if } h_t<h_t^\ast\\
    0, & \text{otherwise}
\end{cases} \,,
\label{eqn:s-hold}
\end{align}
where $\alpha$ is a dimensionless constant. While \cite{Daganzo2009} employs this parameter as part of a real-time control strategy, of which $\alpha$ is a dynamical variable which changes based on the current demand, we will limit our scope of study to static values of $\alpha$. A holding strategy with $\alpha > 1$ over-reacts to its current headway e.g. if $\alpha =2$, a bus which is $1$ minute behind its scheduled headway will hold for $2$ minutes. Conversely, $\alpha < 1$ models an under-reaction to its headway. At the $\alpha = 0$ limit, the bus has \emph{no reaction} to its headway and does not hold for any amount of time. In this case, the system behaves as the control, where there is no intervention in place. On the individual level, holding strategies increase the travelling time of passengers on-board the held bus. However on a system level, the negative feedback would keep the bus system close to the equal-headway state, which would give rise to a lower average travelling time. 
\begin{figure*}[ht]
  \centering
    \begin{subfigure}{\linewidth}
        \centering
        \includegraphics[width=\textwidth]{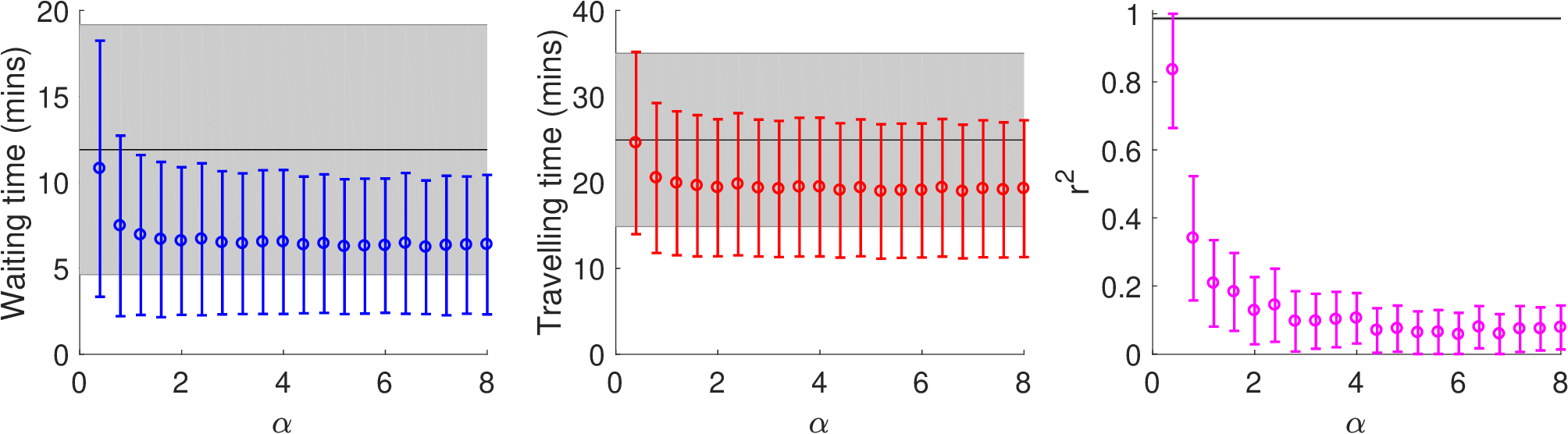}
        \caption{S-holding strategy in lull-same}
        \label{fig:lull_same1}
        \vspace{1.00cm}
    \end{subfigure}
    \begin{subfigure}{\linewidth}
        \centering
        \includegraphics[width=\textwidth]{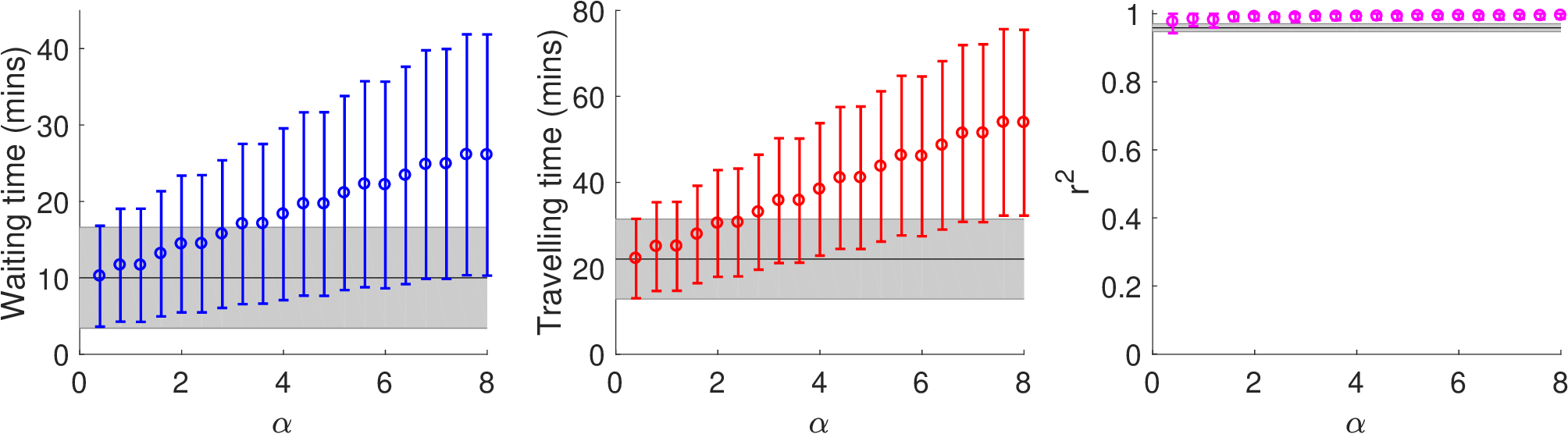}
        \caption{S-holding strategy in busy-same}
        \label{fig:busy_same1}
        \vspace{1.00cm}
    \end{subfigure}
    \caption{Results of the s-holding strategy in the two same-velocity scenarios. Each colored point represents the mean and standard deviation of the measured quantities calculated from 100 realizations of each parameter value. The black line and grey shaded area of each graph denote the mean and standard deviation of the measured quantities from a control case, which are simulation runs of the same parameter values, without any intervention strategies.}
    \label{fig:s-hold}
\end{figure*}

%\emph{Lull phase}
Results of lull-same scenario in Fig. \ref{fig:lull_same1} shows that the s-holding strategy is able to reduce the average waiting and travelling time of the passengers. In the absence of interventions ($\alpha =0$), the $r^2$ shows the buses sustaining a completely bunched state, resulting in a long waiting time of around $12$ minutes per passenger. As $\alpha$ increases, the state of complete bunching decreases with both the waiting and the travelling time, with waiting time decreasing up to 50\% and travelling time decreasing up to 20\%. The difference in savings between waiting time and travelling time can be attributed to the mechanism of the holding strategy. When a bus is being held, the time passengers spend on-board the bus increases. However, this mechanism does not directly increase the waiting time of the passengers at any of the bus stops. In fact, it only decreases them, on average, as the buses are more evenly spread out. 

As the s-holding strategy increases the travelling time of some of the passengers, one might expect that excessive holding - with large $\alpha$ - might result in a decrease in the system efficiency. Interestingly, s-holding performs well over a large range of $\alpha$. By investigating the dynamics of the system at large $\alpha$, we found that excessive holding time also results in a large extension of the bus headway. As s-holding strategy only triggers on buses with small headway, buses which are held with large $\alpha$ gain a large headway allowance, which leads to more infrequent holding. This trade-off between longer but more infrequent holding results in the system maintaining the average travelling time over a range of $\alpha$. This relationship however, fails at extreme values of $\alpha$, where one bus is held for a time longer than the effective period of the other bus. As this value represents a completely impractical strategy, it is not considered in our study. While larger values of $\alpha$ do produce an improvement in the average travelling time, traffic planners do need to consider the practicality of holding a bus for an extended period, especially when it affects the sentiments of the passengers. With that consideration, we recommend the smallest $\alpha$ in which buses do not bunch. 

%\emph{Busy phase}
On the other hand, Fig. \ref{fig:busy_same1} plots the results of the s-holding strategy in the busy phase. From the control cases of $\alpha = 0$, buses with the same velocity exhibit complete bunching at equilibrium, similar to the lull phase. However for $\alpha > 0$, the waiting time of the passengers increases linearly with $\alpha$, representing a complete failure of the s-holding strategy. Furthermore, the comparison between the uncertainty of the $r^2$ between $\alpha = 0$ and $\alpha > 0$ highlights a marginal increase in the degree of bunching for $\alpha > 0$. This suggests that instead of alleviating bus bunching, the s-holding strategy in the busy phase could induce buses to bunch even more. This phenomenon will be further discussed in Sect. \ref{sect:intdiff}, when we look at the effects of s-holding in the busy-different scenario.

\subsubsection*{Intervention: Continuous-time Holding}
In the previous stop-based holding strategy, the time-headway was calculated by time difference utilizing the departure times from the bus stops. Arguably, it is not the only way to define the time-headway of the vehicle. In fact, the concept of time-headway (or any vehicular headway in general) is not well-defined in the literature. In this regard, Ref. \cite{Andres2017} proposed another definition of time-headway as the time taken for a vehicle to reach the current position of the leading vehicle. This was referred to as \emph{continuous-time} time-headway. 

The intricacy of using continuous-time time-headway is that it requires prediction to be made of a vehicle’s trajectory. As such, it has not been used as a viable measurement for most studies on holding strategy. Ref. \cite{Andres2017} tested four different headway prediction methods for its continuous-time holding strategy, finding similar performance result between linear regression and extrapolation, kernel regression and extrapolation, artificial neural networks and autoregressive models. In this study, we propose another headway prediction method for continuous-time holding strategy, which is based on the input parameters (i.e. the empirical distributions) of the EMB model. Excluding the bus stops, the average time taken for a bus to traverse a site can be derived from the inverse of the average velocity at the site. This can be easily calculated from the empirical distributions. At the bus stops, the time spent would be equal to the stoppage time due to passenger alighting and boarding. As the number of passengers at a bus stop is dependent on its headway, by assuming all buses are in an equal time-headway state, i.e. $h_t = H_t/N$, the stoppage time at bus stop $j$ is given by:
\begin{equation}
\tau_j = \frac{\lambda_j}{l}h_t =\frac{\lambda_j L(l+\sum_{j=1}^M \lambda_j)}{l^2 N \bar{v}} \,,
\label{eqn:htmap}
\end{equation}
where we have used $H_t=T_0(1+\sum_{j=1}^M \lambda_j/l)$ and $T_0 = L/\bar{v}$. Note that $T_0$ is the intrinsic period of the bus, $L$ is the length of the route, and $\bar{v}$ is the average intrinsic velocity of the bus deduced from the empirical velocity distributions. Also, recall that $M=12$ bus stops for NTU-SBS.

\begin{figure}[ht]
  \centering
  \includegraphics[width=\linewidth]{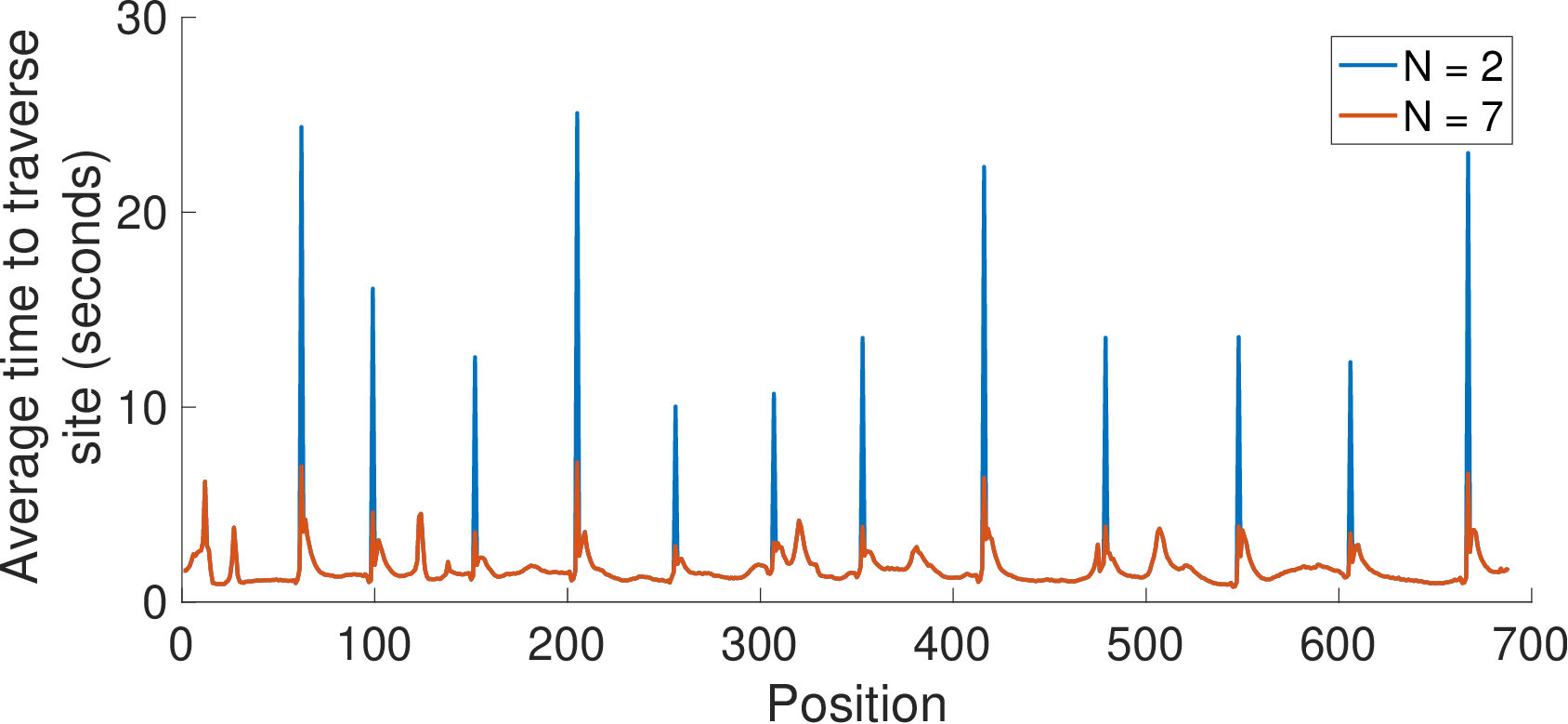}
  \caption{The average time the bus takes to traverse each site of the route based on the number of active buses. The total time taken for a bus to reach its destination position can be derived by taking the total area under the curve between the two points. }
  \label{fig:timeheadwaymap}
\end{figure}
With that, Fig. \ref{fig:timeheadwaymap} plots the time taken to traverse each site for different number of buses in the system. Based on the current position of both the departing bus and its leading bus, Fig. \ref{fig:timeheadwaymap} can be used to determine the time taken for the departing bus to reach the current position of the leading bus. As with s-holding, the resultant headway is used to compute the holding time of the bus based on Eq.~(\ref{eqn:s-hold}). As this strategy uses the continuous-time time-headway, it will be referred to as the continuous-time holding strategy, or c-holding for short.

\begin{figure*}[ht]
  \centering
    \begin{subfigure}{\linewidth}
        \centering
        \includegraphics[width=\textwidth]{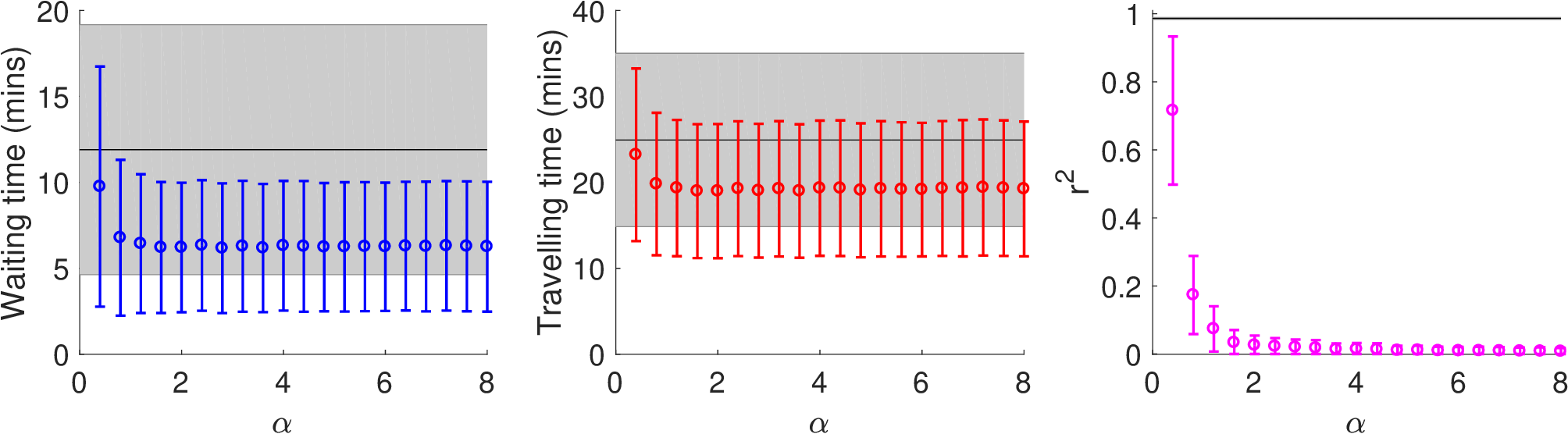}
        \caption{C-holding strategy in lull-same}
        \label{fig:lull_same2}
        \vspace{1.00cm}
    \end{subfigure}
    \begin{subfigure}{\linewidth}
        \centering
        \includegraphics[width=\textwidth]{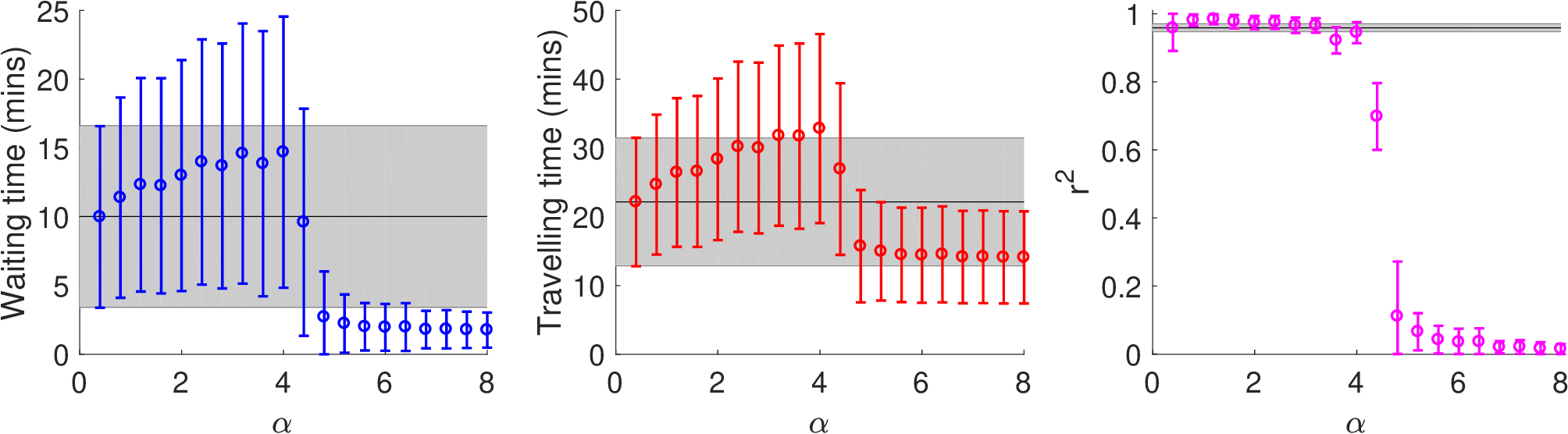}
        \caption{C-holding strategy in busy-same}
        \label{fig:busy_same2}
        \vspace{1.00cm}
    \end{subfigure}
    \caption{Results of the c-holding strategy in the two same-velocity scenarios. Each colored point represents the mean and standard deviation of the measured quantities calculated from 100 realizations of each parameter value. The black line and grey shaded area of each graph denote the mean and standard deviation of the measured quantities from a control case, which are simulation runs of the same parameter values, without any intervention strategies.}
    \label{fig:c-hold}
\end{figure*}

Figure \ref{fig:c-hold} plots the effects of the c-holding strategy on the lull-same and busy-same scenarios. In both scenarios, we see that c-holding performs better than the s-holding strategy. Notably, the biggest improvement over the s-holding strategy is that the c-holding strategy is able to keep the buses unbunched in the busy phase. The major difference between the two holding strategies is that the headway measurement in c-holding factors in the specific dynamics of the bus system (passenger arrival and bus velocity). This suggests that an effective holding strategy has to account for the specific dynamics of the buses, which is absent in the s-holding strategy. This hypothesis will be further analyzed after comparing the effects of both holding strategies in the busy-different scenario in Sect. \ref{sect:intdiff}

\subsubsection*{Intervention: No-boarding}
Unlike holding strategy, which effectively slows down a bus when it draws too near to the leading bus, a no-boarding strategy serves to speed up buses which have their trailing buses too near \cite{Saw2020}, i.e, the backward headway $h^\dagger_d$ (or the trailing bus's $h_d$) is small. As the act of boarding passengers requires buses to stop and dwell at the bus stops, they can be `sped up' by not boarding the passengers at the bus stops. While these buses will still stop for passengers alighting, this strategy reduces the additional dwell time due to passenger boarding, thereby allowing the bus to pull away from the trailing bus. In the ensuing analysis, the concept of `speeding up' the buses by the no-boarding strategy and `slowing down' buses by the holding strategy will be important. 

In a recent work, Ref. \cite{Saw2020} studied the effects of no-boarding with extensive simulations, and provided analytical calculations for the average waiting time under the effects of no-boarding. In that paper, the no-boarding strategy is parameterized by a threshold distance, such that buses stop boarding passengers when their backward distance headway $h^\dagger_d$ falls below a particular threshold $h^\dagger_{d,c}$. For $h^\dagger_{d,c} = 0$, no-boarding never gets activated, allowing it to be used as a control. The authors found an upper-bound in the threshold such that when $h^\dagger_{d,c}$ is greater than the threshold, the waiting and travelling times of the passengers increase drastically. Consider the no-boarding strategy in a system of 2 buses. If $h^\dagger_{d,c} \geq L/2$ with $L$ being the distance of the whole route, boarding will be permanently disabled on at least one of the buses regardless of their configuration. In fact, if the buses are in an equal-headway state, both buses will not be boarding passengers. In general, the upper-bound of the threshold was found to be strictly less than $L/N$. Based on this, we limit the parameter $h^\dagger_{d,c} \in [0, L/N]$ in our simulation studies.

\begin{figure*}[ht]
  \centering
    \begin{subfigure}{\linewidth}
        \centering
        \includegraphics[width=\textwidth]{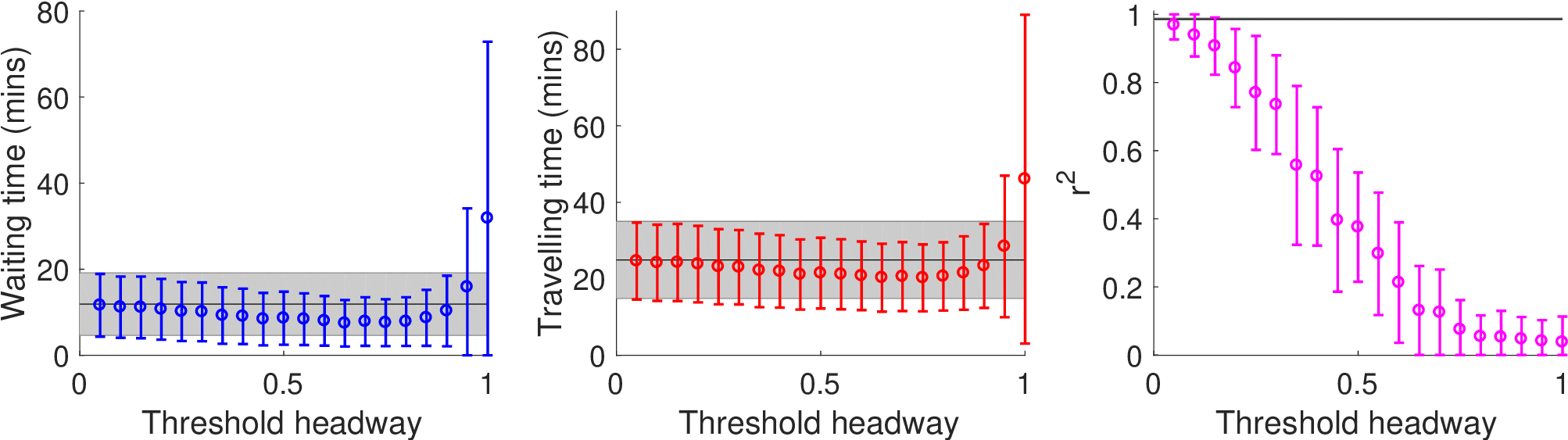}
        \caption{Distance-based no-boarding strategy in lull-same}
        \label{fig:lull_same3}
        \vspace{1.00cm}
    \end{subfigure}
    \begin{subfigure}{\linewidth}
        \centering
        \includegraphics[width=\textwidth]{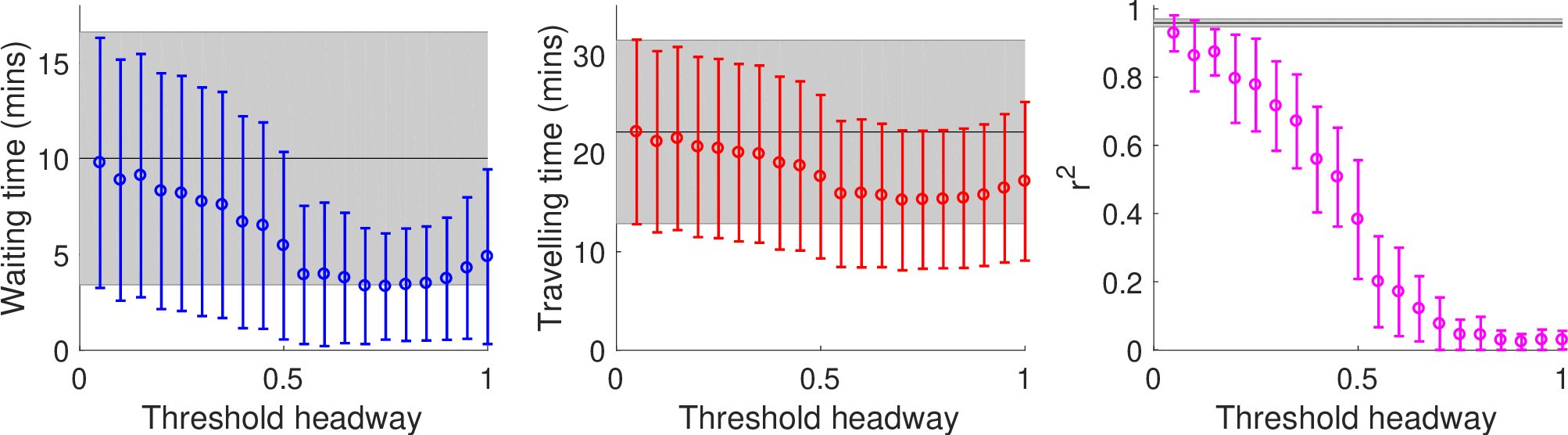}
        \caption{Distance-based no-boarding strategy in busy-same}
        \label{fig:busy_same3}
        \vspace{1.00cm}
    \end{subfigure}
    \caption{Results of the distance-based no-boarding strategy on the two same-velocity scenarios. Each colored point represents the mean and standard deviation of the measured quantities calculated from 100 realizations of each parameter value. The black line and grey shaded area of each graph denote the mean and standard deviation of the measured quantities from a control case, which are simulation runs of the same parameter values, without any intervention strategies.}
    \label{fig:nb}
\end{figure*}

\begin{figure*}[ht]
  \centering
    \begin{subfigure}{\linewidth}
        \centering
        \includegraphics[width=\textwidth]{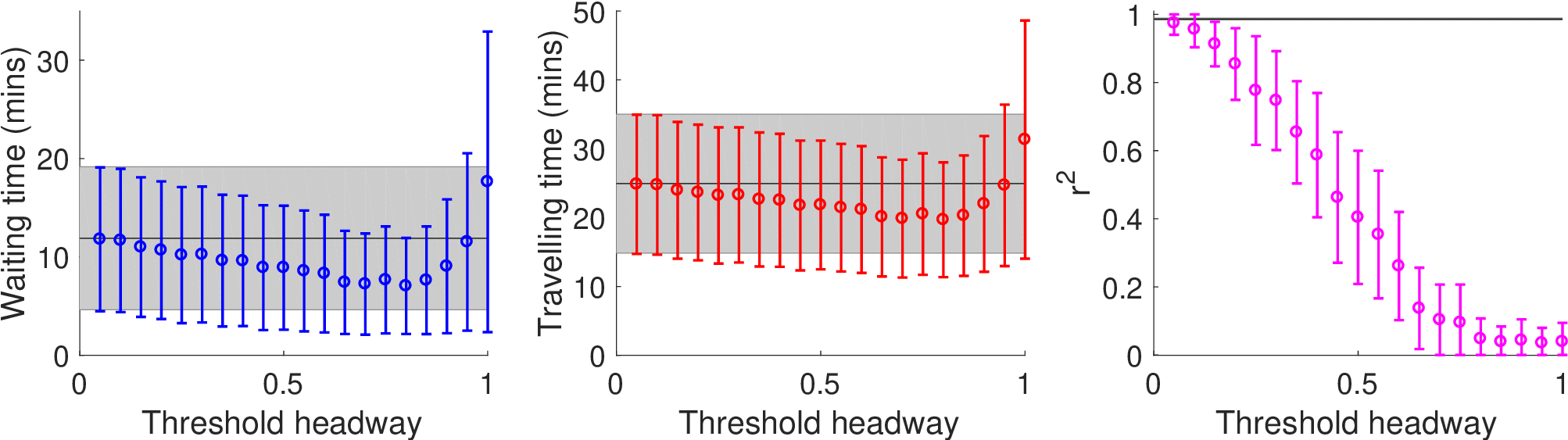}
        \caption{Time-based no-boarding strategy in lull-same}
        \label{fig:lull_same4}
        \vspace{1.00cm}
    \end{subfigure}
    \begin{subfigure}{\linewidth}
        \centering
        \includegraphics[width=\textwidth]{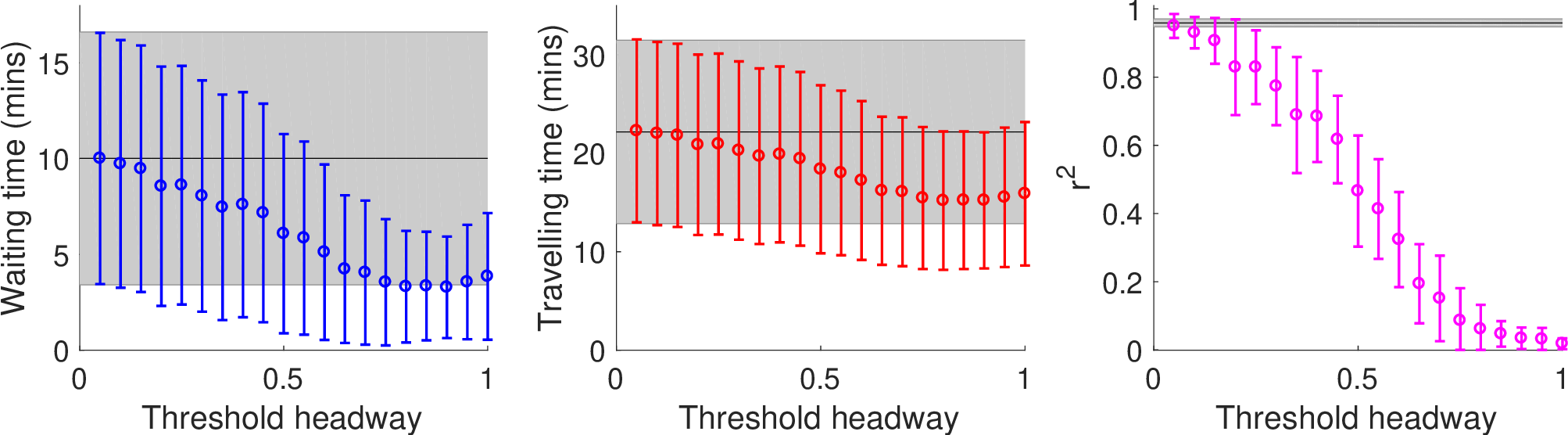}
        \caption{Time-based no-boarding strategy in busy-same}
        \label{fig:busy_same4}
        \vspace{1.00cm}
    \end{subfigure}
    \caption{Results of the time-based no-boarding strategy on the two same-velocity scenarios. Each colored point represents the mean and standard deviation of the measured quantities calculated from 100 realizations of each parameter value. The black line and grey shaded area of each graph denote the mean and standard deviation of the measured quantities from a control case, which are simulation runs of the same parameter values, without any intervention strategies.}
    \label{fig:nbt}
\end{figure*}

As the distance-headway does not take into account the time taken to traverse different segments of the roads (as described by Fig. \ref{fig:vcolormap}), using distance-headway might introduce significant inaccuracy to the no-boarding strategy. A more accurate approach could be to use the time-headway of the buses. Similar to the distance-based approach, the time-based no-boarding strategy will disable boarding on buses that have a backward time-headway $h^\dagger_t$ smaller than $h^\dagger_{t,c}$. The upper-bound is set at $h^\dagger_{t,c} = H_t/N$ where $H_t$ is the period of an average bus. With that, Figs. \ref{fig:nb} and \ref{fig:nbt} plot the simulation results of both the distance-based and time-based no-boarding strategies respectively. 

By comparing Figs. \ref{fig:nb} and \ref{fig:nbt}, no significant difference was found between the distance-based and time-based no-boarding strategy. This refutes our hypothesis that there will be significant inaccuracy brought forth by using distance-headway. In our simulation methodology, the parameter $h^\dagger_{d,c}$ was studied by varying in step sizes of $0.05 L/N$. Based on the relationship between distance and time-headway from Fig. \ref{fig:timeheadwaymap}, the inaccuracy generated by the road heterogeneity accounts for $0.015L/N$, which is less than the simulation step-size. This represents insignificant inaccuracy as compared to the parameter step-size, which leads to the resulting similarity between the two strategies. Based on this, further study of the no-boarding strategy will focus only on the distance-based no-boarding.

%\emph{Lull phase}
Fig. \ref{fig:lull_same3} plots the effects of the no-boarding strategy in the lull-same scenario. Fundamentally, no-boarding is able to keep the buses in a staggered configuration (described by $r^2 \simeq 0$ in Fig. \ref{fig:lull_same3}). Nonetheless, if the threshold headway is too large (such as $h^\dagger_{d,c} = L/N$), no-boarding becomes too frequent and the consequence is a larger waiting time, even as the buses are kept staggered. The minimum waiting time occurs with a threshold headway of $h^\dagger_{d,c} =0.82 L/N$, with a 37\% savings in average waiting time and a 20\% savings in average travelling time.

%\emph{Busy phase}
On the other hand, the busy-same phase in Fig \ref{fig:busy_same3} shows similar dynamics to that of the lull phase, such that the no-boarding strategy is able to improve the waiting and travelling time of the passengers. One interesting feature of the results is that there is an improvement (over the control) in waiting and travelling times of the passengers for $h^\dagger_{d,c} = L/N$. This suggest that the upper-bound of the threshold is larger than the theoretically predicted $L/N$.

The ensuing analysis of the simulated trajectories reveals that there is consistently a certain degree of localized bunching when the theoretical upper-bound is exceeded. As an example, we have Buses 1 and 2 of the 7 buses being locally bunched at angular position $0^\circ$, Buses 3 to 7 taking $60^\circ$, $120^\circ$, $180^\circ$, $240^\circ$, $300^\circ$ respectively. As there are 7 buses, the threshold headway (in degrees) is $360^\circ / 7 = 51.4^\circ$. Since Buses 3 - 7 have backward headway larger than the threshold, they are able to board passengers most of the time, and maintain their headway by executing no-boarding when their trailing buses pulls too close. Due to the presence of stochasticity, Buses 1 and 2 will likely not be in the exact same position. In the case where Bus 1 precedes Bus 2 by a small distance, Bus 1 will have a backward headway of approximately zero and Bus 2 will have a backward headway of $60^\circ$ (with reference to Bus 3). As such, while Bus 1 will not be able to pick up any passengers, Bus 2 will. This causes Bus 1 and Bus 2 to behave as a single bus, thereby resulting in a system where there are 6 active buses, maintaining a mutual headway (in distance) of $L/6$ which is modulated by a no-boarding strategy of threshold headway $L/7$. This results in a more efficient system when compared to the control (no intervention strategies) where all 7 buses will eventually bunch into a single platoon. 

This finding is not a contradiction to the theoretical work presented in Ref. \cite{Saw2020}, which is based on buses having the same constant velocity. In the case where two buses of the same constant velocity are bunched, both buses will have (and constantly maintain) a backward headway of exactly zero. This causes both buses to be in a perpetual state of no-boarding. Consequently, these two buses will catch up with a preceding bus, where the minimized headway causes that bus to also execute no-boarding. The cycle repeats with all other buses until none of the buses are boarding. The existence of stochasticity in our simulation study allows the system to continually exist in a localized bunched state, which allows high-threshold no-boarding strategy to perform better than otherwise predicted. This exemplifies a system where the specificity plays an important role in determining the system dynamics.

\subsubsection*{Intervention: Centralized-pulsing}
In the theory of synchronization, self-oscillators can be synchronized by a periodic pulse \cite{PIK01}. As an example, clocks these days need not be expensively accurate. Instead, a high-accuracy centralized clock can periodically send out electromagnetic waves to less expensive \emph{radio-controlled} clocks to periodically entrain them, such that the latter's accuracy is maintained by being synchronized with the former. This example suggests the idea of ``entraining'' the buses by a periodic force to prevent them from bunching together \cite{Saw2019}.

The new strategy is to subject the buses to a system of periodic driving forces that keep them in a staggered configuration. While the two previous strategies deal with the buses' interaction with passengers, this intervention involves the intrinsic behaviors of the buses. The strategy involves actuating each bus periodically (every $\Delta t_p$ time steps) to either speed up or to slow down, based on their current headway. Specifically, each bus with a backward time-headway greater than the equal time-headway state of $h_t^\ast = H_t/N$ will be forced to slow down, and buses with $h_t < H_t/N$ will be sped up. As with the previous two strategies, this effect acts as a negative feedback for the buses to maintain an equal time-headway state.
In our model, speeding up is carried out by restricting the sampling of velocity from the upper median of the corresponding empirical velocity distribution of the sites the buses are at. Conversely, slowing a bus down involves sampling only velocities below the median of the distribution. 

By varying the parameter $\Delta t_p$ at which this mechanism is applied, we are able to vary the effect of the intervention strategy. To be consistent with the other strategies, $\Delta t_p\rightarrow\infty$ represents a control, in which the buses are never nudged to move faster or slower. The maximum $\Delta t_p = 100$ represents actuating the buses to move above/below its average speed only 1\% of the time. Conversely, $\Delta t_p =1$ models buses being sped up/slowed down \emph{all the time}. The practicality of this will be discussed below. 

\begin{figure*}[ht]
  \centering
    \begin{subfigure}{\linewidth}
        \centering
        \includegraphics[width=\textwidth]{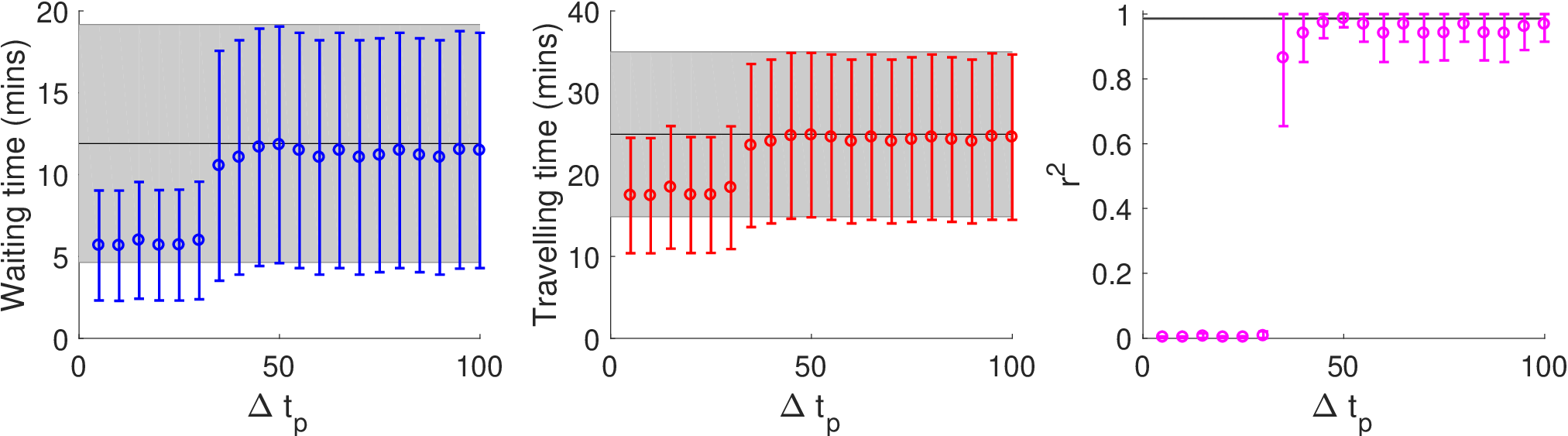}
        \caption{Centralized-pulsing strategy in lull-same}
        \label{fig:lull_same5}
        \vspace{1.00cm}
    \end{subfigure}
    \begin{subfigure}{\linewidth}
        \centering
        \includegraphics[width=\textwidth]{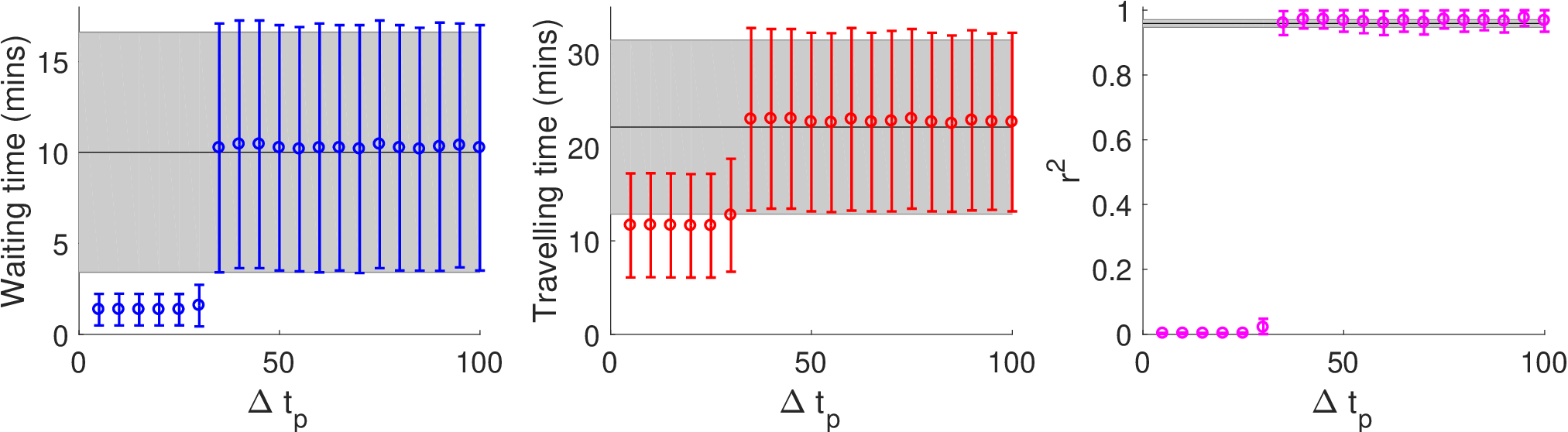}
        \caption{Centralized-pulsing strategy in busy-same}
        \label{fig:busy_same5}
        \vspace{1.00cm}
    \end{subfigure}
    \caption{Results of the centralized-pulsing strategy on the two same-velocity scenarios. Each colored point represents the mean and standard deviation of the measured quantities calculated from 100 realizations of each parameter value. The black line and grey shaded area of each graph denote the mean and standard deviation of the measured quantities from a control case, which are simulation runs of the same parameter values, without any intervention strategies.}
  \label{fig:pulsing}
\end{figure*}

The results presented in Fig. \ref{fig:pulsing} agrees with our intuition that with a higher actuating frequency, the buses stay unbunched. Correspondingly, the interventions with large $\Delta t_p$ have no effect on either the $r^2$ or the waiting/travelling times. As there are similar dynamics between the same-velocity and the different-velocity scenarios, the detailed analysis of the centralized-pulsing strategy will be presented with the results in the next subsection. 

\subsection{Effects of intervention strategies on different-velocities scenarios}
\label{sect:intdiff}
%Before analyzing the effects of the intervention strategies, we will first discuss the dynamics of the buses in the lull-different and busy-different scenarios, in the absence of any interventions.

Reference \cite{Saw2019} discussed that in the lull-different scenario, the buses exhibit periodic bunching. This is the case where the fast bus periodically overtakes the slower bus and, as such, the buses are unable to maintain sustained bunching. A preliminary analysis found that a bus system exhibiting periodic bunching is a relatively efficient system. This can be seen in the simulation results of this section when $\alpha=0$ (e.g. in Fig. \ref{fig:lull_diff1}) where the average waiting time of passengers is $7.4$ minutes compared to $12$ minutes when the bus velocities are the same (in Fig. \ref{fig:lull_same1}). 

On the other hand, unlike the lull-different scenario, the high passenger arrival rate in the busy-different scenario does result in some degree of sustained bunching. However, due to the velocity differences, the dynamics are neither completely sustained bunching nor periodic bunching. The system takes on a complex dynamical state of buses bunching locally in smaller platoons, with the size and number of platoons fluctuating. An important feature is that these ``bus platoons'' are scattered randomly along the bus route. The lower value of $r^2$ ($\alpha = 0$ of Fig. \ref{fig:busy_diff1}) is a result of platoons of bunched buses spanning the bus route and does not imply a complete absence of bus bunching. This is an example where $r^2 = 0$ cannot be interpreted as an equal-headway state when $N>2$. Nonetheless, this complex dynamical state serves as a more efficient state than the completely bunched state, with waiting time averaging at $4.9$ minutes (see Fig. \ref{fig:busy_diff1}) instead of $10$ minutes (see Fig. \ref{fig:busy_same1}). 

\subsubsection*{Intervention: S-holding}

\begin{figure*}[ht]
  \centering
\begin{subfigure}{\linewidth}
        \centering
        \includegraphics[width=\textwidth]{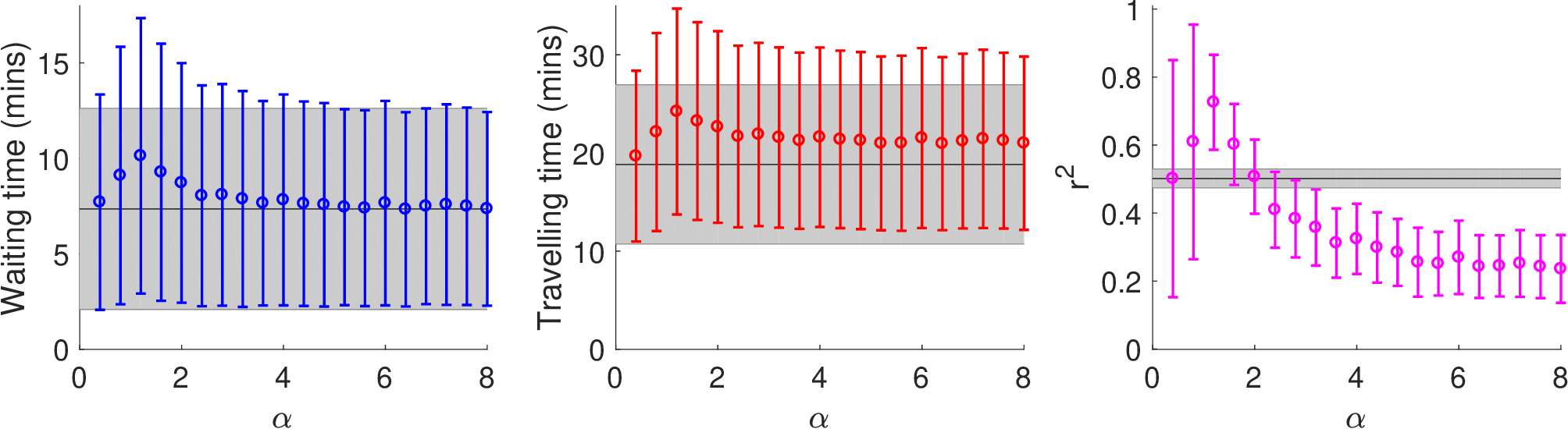}
        \caption{S-holding strategy in lull-different}
        \label{fig:lull_diff1}
        \vspace{1.00cm}
    \end{subfigure}
\begin{subfigure}{\linewidth}
        \centering
        \includegraphics[width=\textwidth]{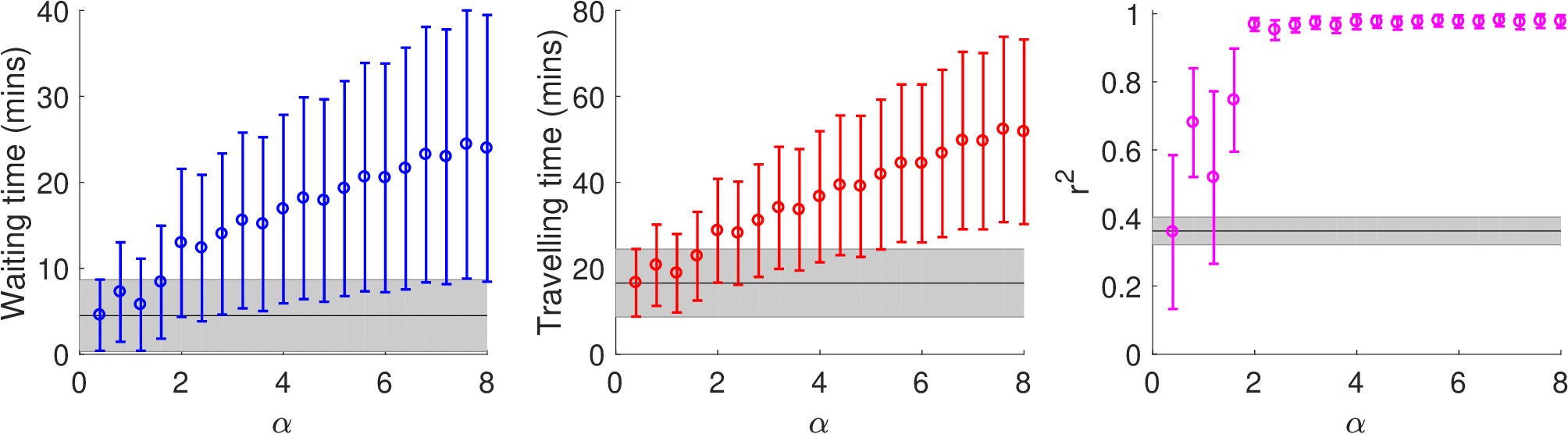}
        \caption{S-holding strategy in busy-different}
        \label{fig:busy_diff1}
        \vspace{1.00cm}
    \end{subfigure}
\caption{Results of simulating the s-holding strategy in the two different-velocity scenarios. Each colored point represents the mean and standard deviation of the measured quantities calculated from 100 realizations of each parameter value. The black line and grey shaded area of each graph denote the mean and standard deviation of the measured quantities from a control case, which are simulation runs of the same parameter values, without any intervention strategies.}
    \label{fig:shold2}
\end{figure*}

%\emph{Lull phase}
From Fig. \ref{fig:lull_diff1}, we see that by invoking the holding strategy on a bus system in the periodic-bunching state (a result of buses having different velocities in the lull phase), the strategy backfires with the control performing better than any $\alpha > 0$. At steady state, only the faster moving bus will have a small headway as it closes towards a slower bus. As such, only the faster buses will be held by the holding strategy. At equilibrium, the holding strategy effectively slows the faster bus by a fixed amount every round, with the amount depending on $\alpha$. With small $\alpha$, the holding strategy slows down the faster buses by a little each round, but not enough to prevent periodic bunching. As there is no effect on the state of bunching, the sole effect of the holding strategy is in increasing the passengers’ waiting and travelling time. The least efficient system exists when $\alpha$ is large enough such that the effective velocity of the faster buses match that of the slower buses. With the buses having the same effective velocity, the bunching mechanism takes over and causes the buses to approach a bunched state ($r^2 \approx 0.8$). This results in the point where waiting and travelling time is at the maximum. The average waiting time of $11$ minutes is comparable to the completely bunched state found in $\alpha =0$ of Fig. \ref{fig:lull_same1}. 

By studying both the $r^2$ and the steady state dynamics, the holding strategy is able to completely nullify the mechanism of periodic bunching with large $\alpha$, resulting in the buses maintaining a staggered configuration. The waiting time in this state is comparable to that of the control, suggesting that the periodic-bunching state and the staggered configuration are both equally efficient in keeping the waiting time of the passengers low. However, the travelling time at large $\alpha$ is marginally worse than the control. This can be attributed to the increase in travelling time of the passengers on-board the held buses. Similar findings were reported in Ref. \cite{Saw2019b} where there was an improvement in waiting time due to holding the faster bus to match the slower bus, albeit at the expense of an increase in average total travelling time.

%\emph{Busy phase}
In the busy phase, as with the results shown in Fig. \ref{fig:busy_same1}, the results of the busy-different scenario also show a linear increase in waiting time, which represents the complete failure of the system. With the no-intervention dynamics of the busy-different scenario averaging an $r^2 = 0.37$, it is apparent that the s-holding strategy with the increased passenger arrival results in a greater degree of bus bunching. While the mechanism of the onset \emph{holding-induced} bunching is unclear, the monotonic relationship between $\alpha$ and waiting time can be explained by the steady state dynamics of the bunched buses. 

The bunching dynamics starts with all 7 buses completely bunched at a single bus stop (also described by the $r^2 \approx 1$). When there are no more boarding passengers, the leading bus (which is not held) leaves the bus stop while the other buses are held by the holding strategy. The leading bus travels towards and stops at the second bus stop to pick up passengers. As all the 6 remaining buses are all bunched into a single platoon which trails the leading bus, the time between this leading bus’s arrival and the previous bus’s departure is very large. This results in a large number of boarding passengers at this second bus stop. As such, the leading bus takes a long time to board the passengers. Throughout our simulation studies, the holding time of the other buses is lower than the time taken for the leading bus to board all the passengers. Therefore, the other 6 buses would have finished holding at the previous bus stop and would then arrive at this second bus stop before the leading bus finished boarding all passengers. With 7 buses boarding concurrently, all the passengers would have boarded a bus and the leading bus would then depart for the next stop. This dynamics of one bus moving to the next stop and the remaining buses following after repeats at every single bus stop. As the speed of the platoon is dependent on the holding time, which is dependent on $\alpha$, the waiting and travelling times of the passengers increase monotonically with $\alpha$. The same set of dynamics occurs in both the busy-same and busy-different scenarios. 

While the rules of any holding strategy would provide a negative feedback to the bus bunching behavior, the s-holding strategy in the busy phase results in a peculiar \emph{holding-induced} bunching. Between the governing equation of the s-holding strategy and the simplistic scheduled headway, we conjecture that the holding-induced bunching can be attributed to its over-simplistic nature which does not take into account important attributes of the bus system. In the next part, we present a more dynamical holding strategy which takes passenger arrival rate and bus velocity into consideration. We will demonstrate that not only it achieves a more efficient bus system but also elaborate on how these attributes would interact with a holding strategy such that a failure to account for them could result in the holding-induced bunching.

\subsubsection*{Intervention: C-holding}
\begin{figure*}[ht]
  \centering
\begin{subfigure}{\linewidth}
        \centering
        \includegraphics[width=\textwidth]{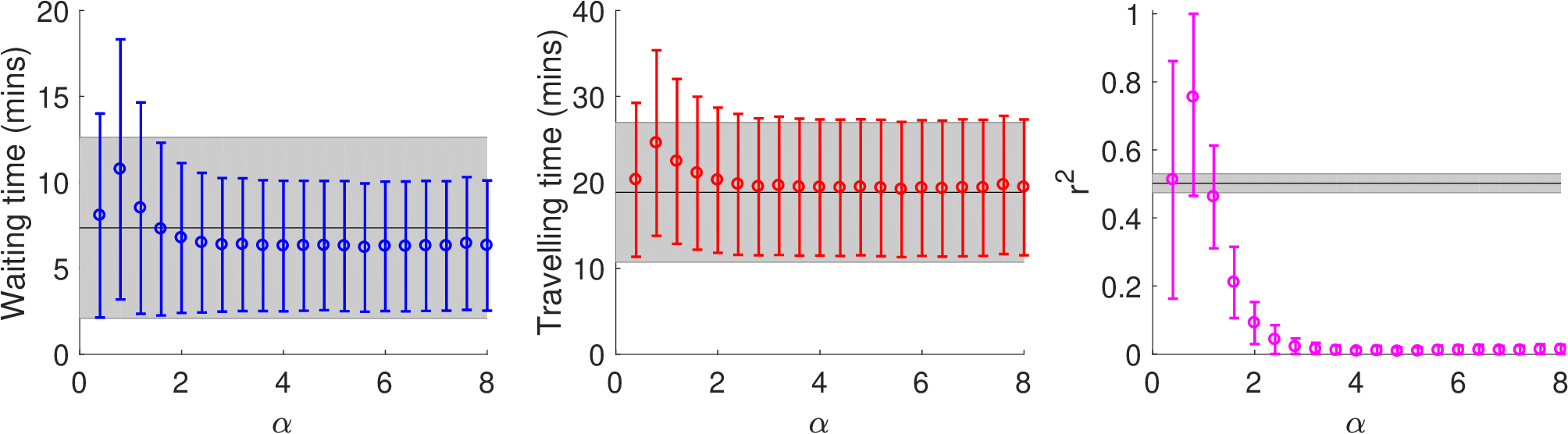}
        \caption{C-holding strategy in lull-different}
        \label{fig:lull_diff2}
    \end{subfigure}
\begin{subfigure}{\linewidth}
        \centering
        \includegraphics[width=\textwidth]{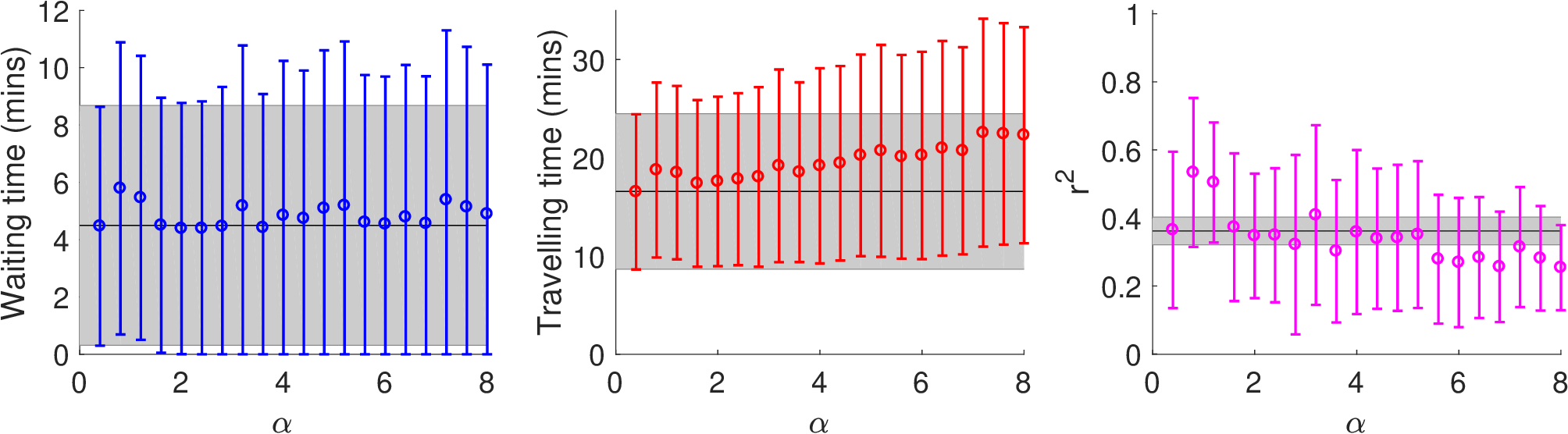}
        \caption{C-holding strategy in busy-different}
        \label{fig:busy_diff2}
    \end{subfigure}
    \caption{Results of the c-holding strategy in the two different-velocity scenarios. Each colored point represents the mean and standard deviation of the measured quantities calculated from 100 realizations of each parameter value. The black line and grey shaded area of each graph denote the mean and standard deviation of the measured quantities from a control case, which are simulation runs of the same parameter values, without any intervention strategies.}
    \label{}
\end{figure*}

In the lull-different case, the dynamics at low values of $\alpha < 1$ is the same as s-holding, in which the c-holding strategy does not keep the buses in the staggered configuration. Unlike the s-holding strategy, Fig. \ref{fig:lull_diff2} shows a marginal improvement in the efficiency at the range of $\alpha > 2$, with a 14\% reduction in waiting time, which is consistent with similar works found in Ref. \cite{Yu2009,Berrebi2018,Saw2019b}.

%As discussed earlier, the \wl{periodic bunching dynamics exhibited by buses in the busy-different scenario leads to an improved efficiency over the busy-same scenario as buses are kept far from complete bunching.}

For the busy-different scenario with the c-holding strategy, the $r^2$ in Fig. \ref{fig:busy_diff2} suggests that the buses are kept at a lesser degree of complete bunching. However, this slight improvement in the state of bunching does not result in an improvement in the system, with average travelling time increasing slightly as $\alpha$ increases. While this implies an improvement over the s-holding strategy, the drop in efficiency suggests that the c-holding strategy does not quite achieve its function when the buses have different intrinsic velocities in a busy phase. 

%\emph{Improvement over the s-holding strategy}
The improvement of the c-holding strategy over s-holding can be attributed to the over-simplistic approach of the former. As the stop-based time-headway is measured as the difference of two successive buses’ departure time, it only ensures that the buses leave each bus stop at or after the scheduled headway. However, it does not consider how headway changes after the holding time. There are two mechanisms which can cause the headway of the buses to change: one involving the bus velocity and the other involving passenger arrival. 

As discussed, when there is a difference in bus velocities, the strategy only holds the bus with the high velocity. Being the faster bus, it will always catch up with its leading bus after holding. The rate at which it catches up depends on the velocity difference between the two buses. To keep the buses exactly at the equal-headway state, a large velocity difference requires a longer holding time than a small velocity difference. By considering the velocities of the buses, the c-holding strategy is able to keep the buses very close to the equal-headway state, suggested by the $r^2$ at large $\alpha$ in Fig. \ref{fig:lull_diff2}. In contrast, the simplistic s-holding strategy in Fig. \ref{fig:lull_diff1} cannot.

At the same time, a holding strategy possesses an inherent self-accentuating effect which is dependent on the passenger arrival rate. Holding a bus extends its distance from its leading bus, effectively increasing its time-headway. An increased time-headway results in additional passenger arrival at the next stop, which in turn leads to a longer stoppage time. As with the velocity considerations, to keep the buses exactly at the equal-headway state requires a shorter holding time when the passenger arrival rate is high and a longer holding time when the arrival rate is low. As the continuous-time measurement of the c-holding strategy takes into account the velocities of the buses and the different arrival rates at different bus stops, it is able to consistently adjust towards an optimal holding time for the buses to converge to the equal-headway state. Conversely, as s-holding does not account for the effects of bus velocity and arrival rates, it is ineffective in maintaining the buses at an equal-headway state, which ultimately results in the buses bunching together.

\subsubsection*{Intervention: No-boarding}
\begin{figure*}[ht]
  \centering
\begin{subfigure}{\linewidth}
        \centering
        \includegraphics[width=\textwidth]{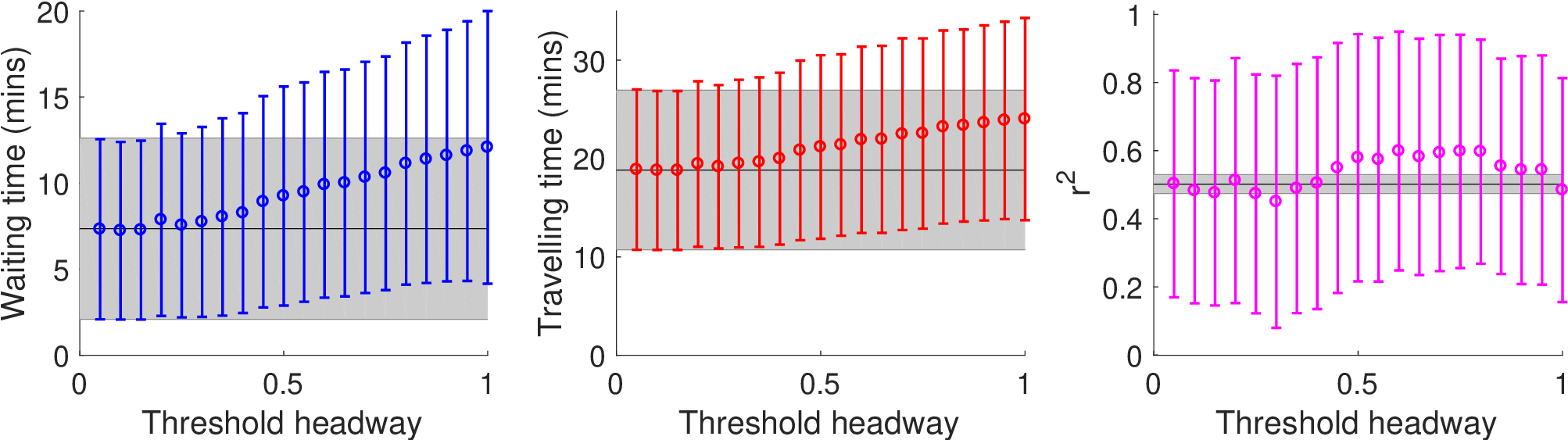}
        \caption{Distance-based no-boarding strategy in lull-different}
        \label{fig:lull_diff3}
    \end{subfigure}
\begin{subfigure}{\linewidth}
        \centering
        \includegraphics[width=\textwidth]{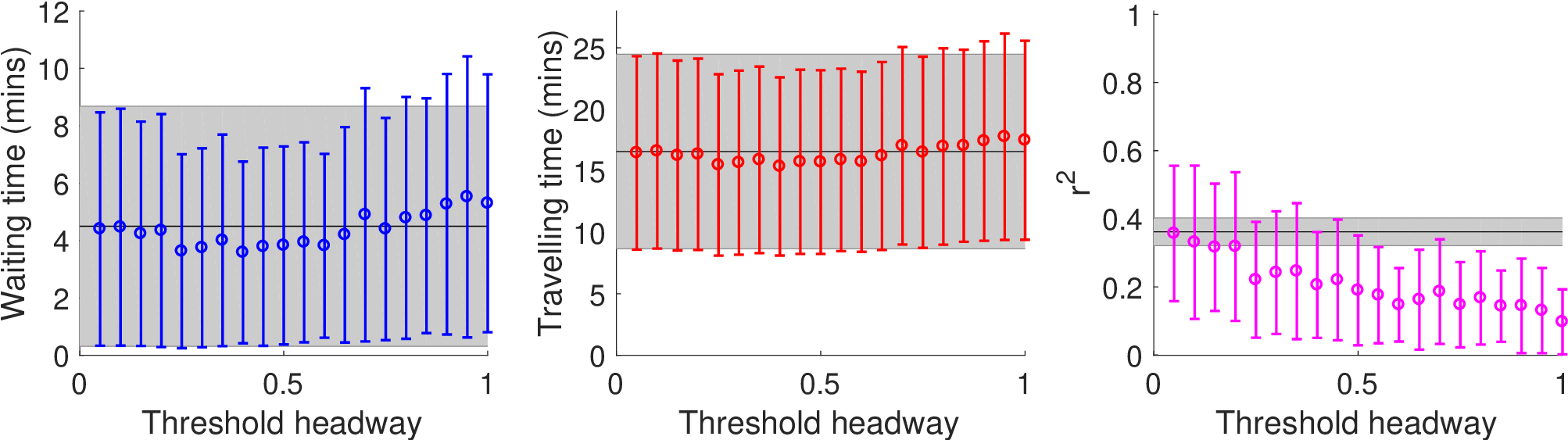}
        \caption{Distance-based no-boarding strategy in busy-different}
        \label{fig:busy_diff3}
    \end{subfigure}
    \caption{Results of the no-boarding strategy in the two different-velocity scenarios. Each colored point represents the mean and standard deviation of the measured quantities calculated from 100 realizations of each parameter value. The black line and grey shaded area of each graph denote the mean and standard deviation of the measured quantities from a control case, which are simulation runs of the same parameter values, without any intervention strategies.}
    \label{fig:nb2}
\end{figure*}

%\emph{Lull phase}
As with the holding strategies, the no-boarding strategy backfires in the lull-different scenario. However, there is a fundamental difference between the holding and no-boarding strategies. In the holding strategy, the headway of the buses are modulated by `slowing-down' the buses through holding. The parameter $\alpha$ determines how long each bus is held, which effectively scales the amount a bus is slowed down by the strategy. In contrast, the parameter $h^\dagger_{d,c}$ only affects the point at which no-boarding is invoked, but does not affect how much a bus is `sped-up'. In the no-boarding strategy, the amount buses are sped-up is dependent on the passenger arrival rate, i.e, the higher the number of passengers \emph{not picked up}, the faster a bus will be \emph{relative to the trailing bus}. Therefore, across the parameter range, the no-boarding strategy behaves like a holding strategy of low $\alpha$ such that the strategy increases the period of the periodic bunching, but does not modulate the buses enough to achieve a staggered configuration (see Fig. \ref{fig:lull_diff3}). This is seen from the equilibrium dynamics of the slower bus being periodically overtaken by the faster bus, with no-boarding triggering when the latter is within its threshold headway. As the only change in the dynamics is the frequency of no-boarding taken by the slower bus, the waiting and travelling times of the passengers increase monotonically with $h^\dagger_{d,i}$.

%\emph{Busy phase}
The results of no-boarding in the busy-different scenario in Fig. \ref{fig:busy_diff3} show marginal improvement in the waiting time of the passengers. This is consistent with the theory presented in \cite{Saw2020} that a higher number of passenger arrival would result in greater degree of speed modulation required to keep the buses unbunched. While our simulation study only captures a marginal improvement in waiting time, Ref. \cite{Saw2020} shows significant savings in waiting time when the passenger arrival rates are even higher (average $k=0.063$, where $k=\frac{\lambda}{l}$). This result is consistent with our analysis and supports that no-boarding is a viable strategy for a very busy bus system, even with buses travelling at different velocities.

\subsubsection*{Intervention: Centralized-pulsing}
\begin{figure*}[ht]
  \centering
\begin{subfigure}{\linewidth}
        \centering
        \includegraphics[width=\textwidth]{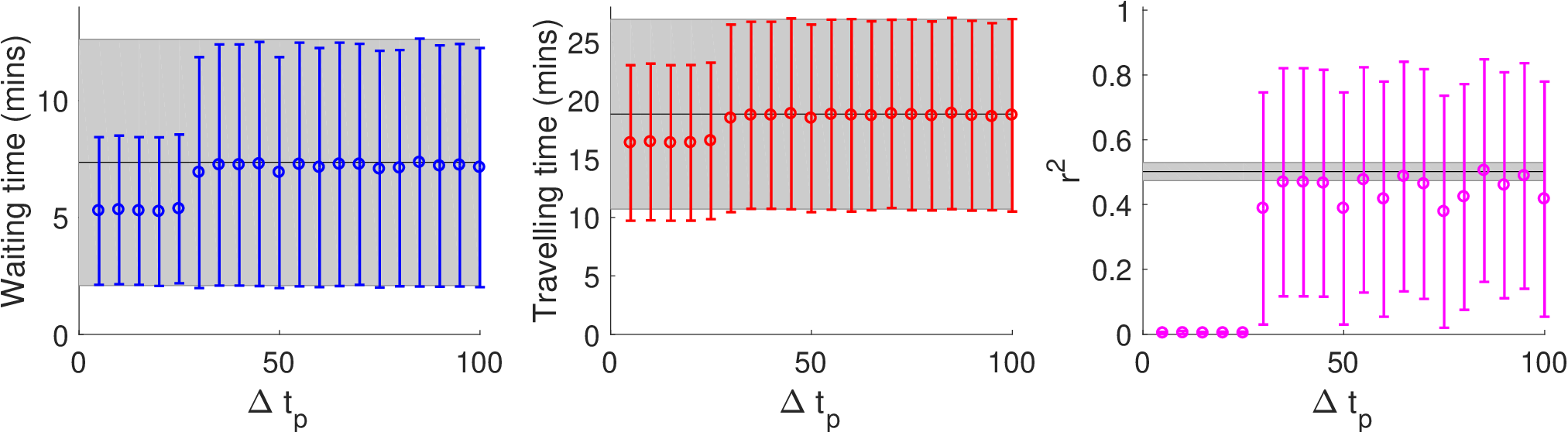}
        \caption{Centralized-pulsing strategy in lull-different}
        \label{fig:lull_diff5}
    \end{subfigure}
\begin{subfigure}{\linewidth}
        \centering
        \includegraphics[width=\textwidth]{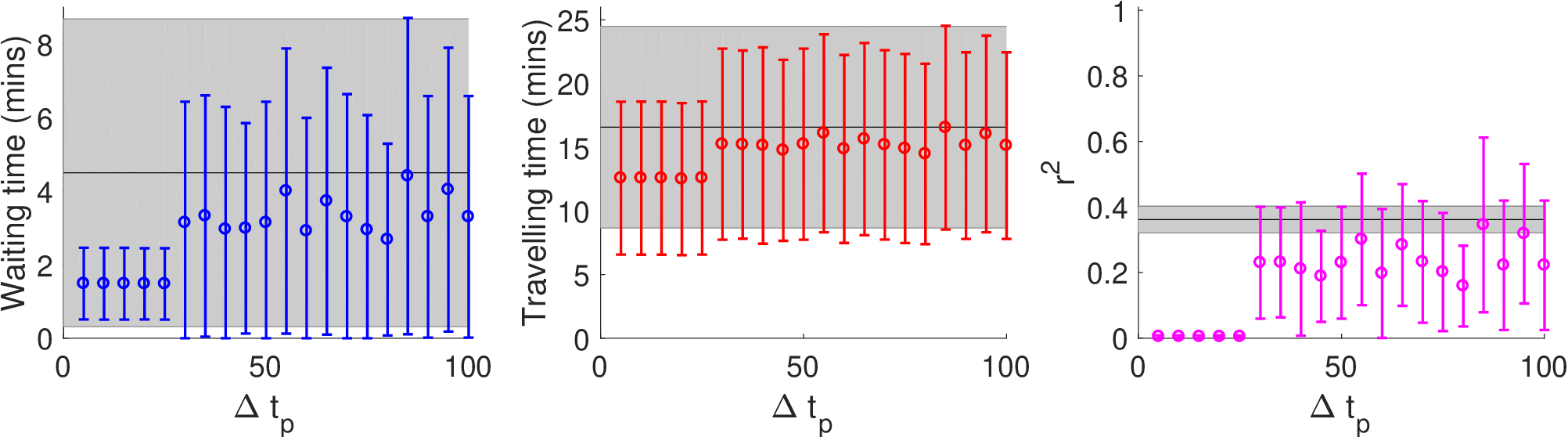}
        \caption{Centralized-pulsing strategy in busy-different}
        \label{fig:busy_diff5}
    \end{subfigure}
    \caption{Results of the centralized-pulsing strategy in the two different-velocity scenarios. Each colored point represents the mean and standard deviation of the measured quantities calculated from 100 realizations of each parameter value. The black line and grey shaded area of each graph denote the mean and standard deviation of the measured quantities from a control case, which are simulation runs of the same parameter values, without any intervention strategies.}
    \label{fig:pulsing2}
\end{figure*}

Comparing Figs. \ref{fig:pulsing} and \ref{fig:pulsing2}, it can be seen that the effects of centralized-pulsing across all four scenarios are very similar. A potentially important finding is the existence of a sharp transition between a completely unbunched and the completely bunched/periodically bunched state, which takes place consistently at $\Delta t_p \approx 30$.

A further analysis was done by studying the system for $\Delta t_p = 30^+$. As discussed, our methodology involves initializing the system in the equal-headway state and measuring the system only in steady state. By investigating the transient state in the lull-same scenario, it is found that centralized-pulsing with $\Delta t_p = 30^+$ maintains the initial equal-headway state for a long time before a sharp transition to the bunched state. This bunching persists in long time limit. This set of dynamics resembles physical systems which exhibit 1st order phase transition, where the system exhibits metastability within a particular range of parameter values \cite{Binder1975,privman_1997,White1989,Schadschneider2002,Pottmeier2003}. At this range, the system could exist in the metastable phase for a long time. However, any perturbations would result in the system undergoing a phase transition into the stable phase. In this case, the equal-headway state is analogous to the metastable state and the bunched state is the stable state.

While centralized-pulsing appears to be a very promising intervention strategy, the practicality of implementation has to be considered carefully. As $\Delta t_p$ is analogous to the `inter-actuation time', $\Delta t_p =1$ models buses consistently moving strictly above or below its average speed based on its current headway. For a bus with slow intrinsic velocity, it has to move above its average velocity \emph{at all times}. Based on the road or traffic conditions, this may not always be possible. In fact, the issue with actuating frequency relates back to the theory of the entrainment of self-oscillators. Discussed in \cite{PIK01}, the onset of synchronisation by a periodic driving force is dependent on both the period and the magnitude of the driving force. In this simulation study, the frequency is set by $\Delta t_p$ and the `magnitude of the force’ is modelled by the velocity selection rule. However, these values might not correspond to the real world. In fact, a reasonable nudging frequency and `magnitude of the force' would not only be driver-dependent, but also spatially dependent, i.e. there will be segments of the road that the driver can choose to drive a little faster or slower. To gather sufficient information to refine our implementation into a realistic and practical centralized-pulsing strategy, an experimental study measuring the bus velocities when prompting drivers to speed up and slow down should be carried out.

\subsection{Comparison of the various intervention strategies}

\begin{table*}\centering
\ra{1.3}
\begin{tabular}{@{}rcrrcrrcrrcrr@{}}\toprule
 & \phantom{abc} & \multicolumn{2}{c}{\textbf{Lull-same (\%)}} & \phantom{abc} & \multicolumn{2}{c}{\textbf{Busy-same (\%)}} & \phantom{abc} & \multicolumn{2}{c}{\textbf{Lull-diff (\%)}} & \phantom{abc} & \multicolumn{2}{c}{\textbf{Busy-diff (\%)}} \\
\textbf{Strategy} && wait & travel && wait & travel && wait & travel && wait & travel \\ 
\midrule
\hline
S-hold && 45.8 & 23.2 && - & - && - & - && - & - \\
C-hold && 48.3 & 24.0 && 82.5 & 34.5 && 14.9 & 0.0 && - & - \\
No-board (dist) && 36.7 & 18.8 && 67.0 & 30.5 && - & - && 4.1 & 0.5 \\
No-board (time) && 40.5 & 20.5 && 66.9 & 31.1 && - & - && 1.6 & 0.0 \\
Pulsing && 53.4 & 59.6 && 86.4 & 46.8 && 28.4 & 10.3 && 63.3 & 26.3 \\
\end{tabular}
\caption{Fractional savings in waiting time and travelling time (respectively in two columns) taken in the four test scenarios, under the four different strategies. The waiting and travelling times of each strategy is represented by its best performing parameter and a null entry represents the control case outperforming the strategy at all parameter values. The fractional savings is given by the fractional difference between the average waiting/travelling time of the strategy and the control case.}
\label{table:int}
\end{table*}

Having modelled the four different intervention strategies, Table \ref{table:int} tabulates the effectiveness of each strategy by the fractional savings in the waiting time and travelling time over a control. Similar works in literature have found savings in waiting times ranging from 5\% to 80\% \cite{Abkowitz1986,McLeod1998,Eberlein1999,Zolfaghari2004,Milla2012,Petit2018,Saw2020,Saw2019b}, which are in agreement with our findings. Among the three classes of strategies, a recurring finding is that centralized-pulsing stands out as the best-performing strategy. With this happening over all scenarios, the comparison between the holding and no-boarding strategies will first be presented.

In the lull phase, both strategies generally only work when the buses have similar intrinsic velocities. While both holding strategies outperform the no-boarding strategy in the lull-same scenario, this marginal out-performance (compared against the uncertainty of the measurement) does not represent any significant benefit in choosing holding strategies over no-boarding for this scenario. In the lull-different scenario, none of the strategies (except centralized-pulsing) provide significant improvement in the system, with c-holding strategy performing marginally better than the control case. By comparing the control case of the lull-same and lull-different scenarios, having buses with different velocities produce an average waiting and travelling time comparable to the no-boarding strategy. This implies that, unlike sustained bunching, periodic bunching is relatively efficient for a bus system. Conveniently, forcing buses to travel with different velocities during lull phase could be another intervention strategy to be undertaken by the bus operators. Alternatively, as smaller buses have a tendency to move faster than larger buses, bus operators could also employ buses of different sizes to increase the tendency of periodic bunching. 

In the busy phase, the holding strategies backfire in most cases. While c-holding could reduce bunching in the busy-same scenario, it requires a relatively long holding time (implied by the large $\alpha$), which would affect its practicality. In contrast, the effectiveness of the no-boarding strategy increases with the number of passengers; in the busy-same scenario, waiting time is reduced by 67\%. While we find the no-boarding strategy to be less effective where there is a large variation in buses' velocity, it was discussed in Ref. \cite{Saw2020} that improvement in efficiency in a busy-different scenario occurs when the passenger arrival rate is much higher. This shows that the no-boarding strategy is potentially an effective strategy in the busy phases.

\section{Conclusion}
\label{sectt:conclude}

In this paper, we propose the Empirically-based, Monte Carlo Bus-network (EMB) model, which is an empirical agent-based model designed as a test bed for potential intervention strategies to reduce bus bunching. Through that, we have studied three classes of intervention strategies. It was found that traditional intervention strategies (holding and no-boarding) only perform well under specific scenarios. Nonetheless, the no-boarding strategy seems to be viable when the passenger arrival rate is high. Of the five strategies that was studied, the centralized-pulsing strategy seems to be the most promising strategy as it was found to be effective across all the different scenarios.

%Further studies are proposed to ascertain its practicality as an implementable strategy. 

%The fully parameterized EMB model is used to study three classes of intervention strategies: holding strategy, no-boarding strategy and a novel centralized-pulsing strategy. It was found that the holding and no-boarding strategies are only effective in specific scenarios, primarily when the velocity of the buses are similar. In contrast, the simulation study found that the centralized-pulsing strategy seems to be effective across a wide range of scenarios, suggesting it as a comprehensive solution for the bus bunching problem. 

As our theoretical studies have found potential solution strategies which would improve their respective systems, future work should strive to bring the solution from a simulation environment to reality. While applications of the EMB model have been presented exclusively on the NTU-SBS, the formulation is generic and can be applied to any other bus system. There are two benefits of applying the modelling framework to other empirical systems. As we have shown, modelling an empirical system using the EMB model allows the users to study and gain insights about the system, facilitating the development of intervention strategies. More importantly, applying the modelling framework to other empirical bus systems can facilitate the continual development of the EMB model. As an example, the interaction between buses and the other cars on the road have not been considered in the current formulation, due to the apparent infrequency of vehicular traffic in NTU's campus. However, if the EMB model is used to model a busy bus route, e.g. one which runs through the city center, the interactions between buses and vehicles become highly significant in determining the dynamics of the bus system. With added features, the EMB model may yield new insights into the modelling framework and even the phenomenon of bus bunching in general.

%While centralized-pulsing was found to be effective across the different scenarios, future work is required to validate the results using experimental data. Here, we propose a future experimental direction. By determining the highest possible frequency of actuating the buses and also the amount of velocity changes when the buses are actuated, additional simulation studies should be performed to ensure the efficacy of the strategy in real-world scenarios. Following which, the technology required for a small-scale implementation in the NTU-SBS is easily available: an algorithm which uses the real-time positions of the buses to determine if a bus should speed up or slow down at the actuation time. With this algorithm built in a mobile app, the cost of even a full-scale implementation would be minimal.

\section*{Acknowledgments}
	This work was supported by the Joint WASP/NTU Programme (Project No. M4082189) and the DSAIR@NTU Grant (Project No. M4082418).

\appendix
\section{Data filtering}
\label{apx:filtering}
\begin{table}[h!]
\centering
\begin{tabular}{||c c c c||} 
 \hline
 Date/time & \thead{Vehicle Registration\\Number} & Latitude ($^\circ$N) & Longitude ($^\circ$E)\\ [0.5ex] 
 \hline\hline
Sep 14 07:54 & PC4964L & 1.353176 & 103.681695 \\ 
Sep 14 07:54 & PC3087A & 1.336900 & 103.677067 \\
Sep 14 07:54 & PC4964L & 1.353176 & 103.681695 \\ 
Sep 14 07:54 & PC3087A & 1.335191 & 103.676586 \\
 ... & ... & ... & ... \\ [1ex] 
 \hline
\end{tabular}
\caption{Raw data collected from https://baseride.com/maps/public/ntu/. The latitude and longitude make up $\tilde{x}_{i,t}$.}
\label{table:rawdatatable}
\end{table}

Table. \ref{table:rawdatatable} illustrates the tabulated raw data collected from \url{https://baseride.com/maps/public/ntu/}. This includes a date/time stamp, vehicle registration number, and the longitude/latitude of each bus as it moves along a designated route. While the data only updates every 9-12s, the script was set to record the data in a 3s interval. This redundancy results in about 60\% of repeated data which was duly processed and removed appropriately as indicated below. The 2-D latitude and longitude of bus $i$ at time $t$ is defined as $\tilde{x}_{i,t}$. As the buses only move along a fixed periodic route, the positions will be mapped onto a discretized 1-D array, where $x_{i,t}$ is an integer which describes the position of the bus along the route. The following paragraphs will describe the methodology to transform $\tilde{x}_{i,t} \rightarrow x_{i,t}$, and the subsequent filtering applied to the data.

The lower limit of spatial resolution was first determined by calculating the minimum $\delta \tilde{x}_{i,t+1}$, where $\delta \tilde{x}_{i,t+1} = || \tilde{x}_{i,t+1}-\tilde{x}_{i,t}||$ for $\tilde{x}_{i,t+1} \neq \tilde{x}_{i,t}$, from all the individual trajectories throughout the sampling period. The spatial resolution of the GPS data was determined by this minimum $\delta \tilde{x}_{i,t+1}$, which is $7.9$m in length. Since a spatial resolution of $7.5$m is typically used in discrete-space traffic models \cite{Nagel1992,Shou-Xin2008,Kerner2014} to reflect the length of an average vehicle and the interaction distance between vehicles, we employ the spatial resolution of $7.5$m (instead of the empirically derived $7.9$m) to allow for future modularity of the simulation, specifically when the simulation is expanded to include the effects of inter-vehicle interactions.

\begin{figure}[h!]
  \centering
  \includegraphics[width=0.7\linewidth]{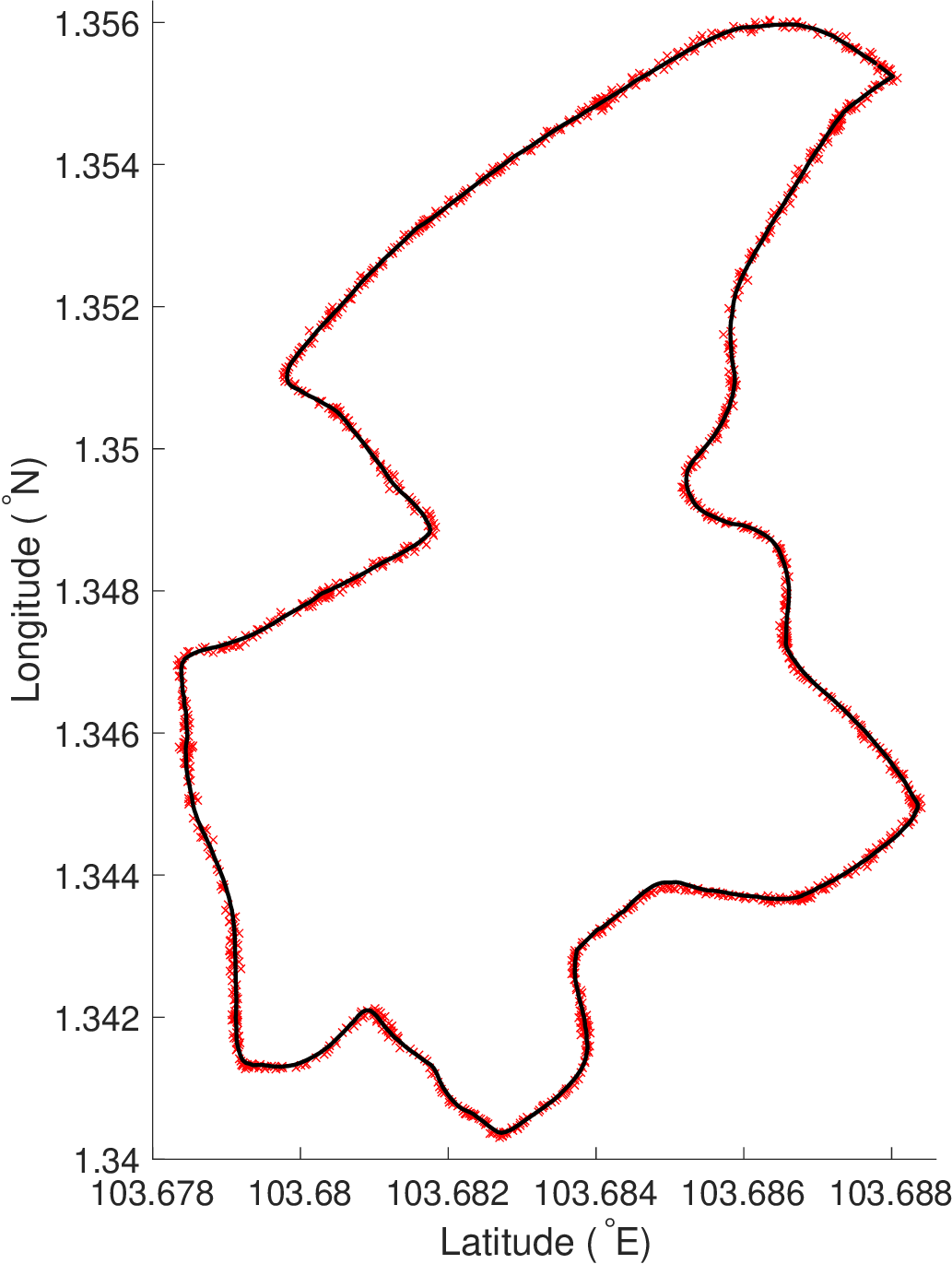}
  \caption{Scatter plot of all $\tilde{x}_{i,t+1}$, with a fitted line representing the average route taken by the buses.}
  \label{fig:NTUscatter}
\end{figure}

All the spatial coordinates collected in the sample period are plotted in a scatter, as illustrated in Fig. \ref{fig:NTUscatter}. A smooth line fitted along the scattered coordinates traces out the bus route. This line is then segmented at regular intervals equivalent to the designated spatial resolution of $7.5$m, resulting in a path made up of 688 evenly distributed points. These points are used to transform the 2-D spatial coordinates, $\tilde{x}_{i,t}$ into a 1-D $x_{i,t}$ path. % ($\tilde{X}_{j}, where $j\in \{1,688\}$). These points are used to transform the collected 2-D into by defining $x_{i,t} = j_0$ if $|\tilde{X}_{j_0} - \tilde{x}_{i,t}| \leq |\tilde{X}_{j} - \tilde{x}_{i,t}|,j \in \{1,688\}$ {\color{red} What is the logic of this mathematical expression?}.

\begin{figure}[h!]
  \centering
  \includegraphics[width=\linewidth]{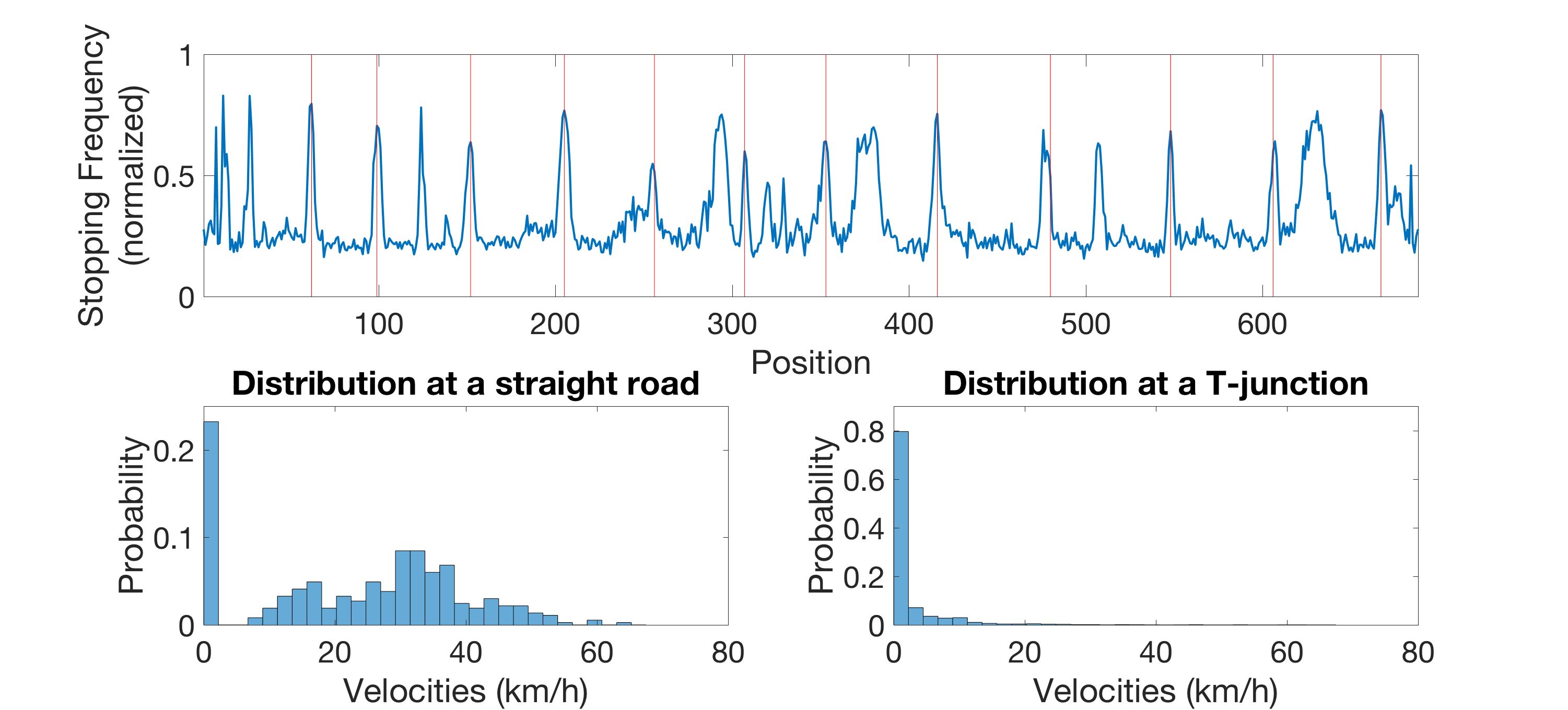}
  \caption{Graphs of normalized stopping frequency i.e. probability of bus stopping at a each position along the bus route, and velocity distributions of two different positions on the road. These graph are derived from the raw data collected over 13 weeks. Red lines on the graph are the sites where there are bus stops.}
  \label{fig:unfilteredvelocities}
\end{figure}

Preliminary studies on the collected data found inaccuracy in the tick-by-tick update of the bus positions, even when taken at 12s intervals. Fig. \ref{fig:unfilteredvelocities} plots the occurrence of stoppages, i.e, $\delta x = 0$, where $\delta x_{i,t} = ||x_{i,t+1} - x_{i,t}||$, along the bus route. While the probability of stoppage peaking at road junctions and bus stops are as expected, the median probability of stoppages is $0.2$. This value corresponds to buses being stationary 20\% of the time, even when traversing a seemingly uninterrupted road. Empirical observation during field work found that this does not reflect the average motion of buses. Further examination on the velocity distribution at single sites (as observed in Fig. \ref{fig:unfilteredvelocities}) also shows the unrealistic propensity of zero-velocities occurring in the middle of a long stretch of road. A plausible explanation for this artefact is that the positioning data sent by the buses was interrupted for a significant period of time while the bus was in motion. As such, there is a possibility that the collected bus positions may not reflect the actual tick-by-tick positions of the buses.

\begin{figure}[h!]
     \centering
     \begin{subfigure}{.48\linewidth}
         \centering
         \includegraphics[width=\textwidth]{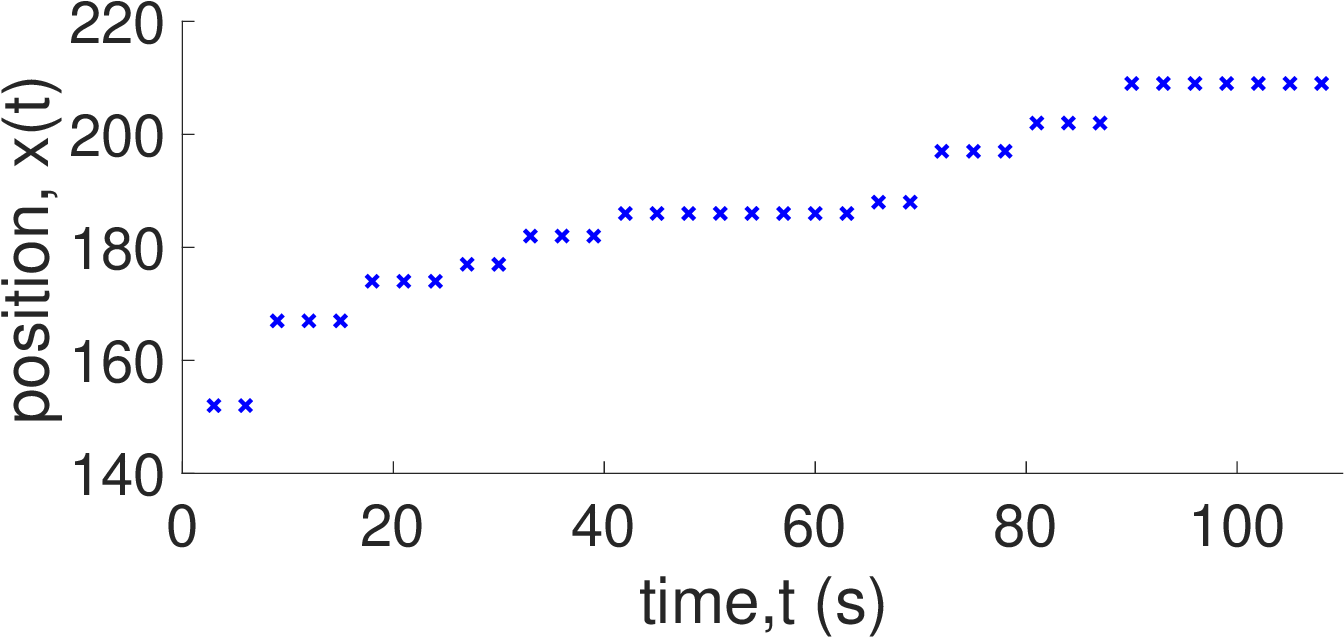}
         \caption{Trajectory, $x(t)$ of a single bus, collected at a time interval of 3 seconds.}
         \label{fig:inter1}
     \end{subfigure}
     \begin{subfigure}{.48\linewidth}
         \centering
         \includegraphics[width=\textwidth]{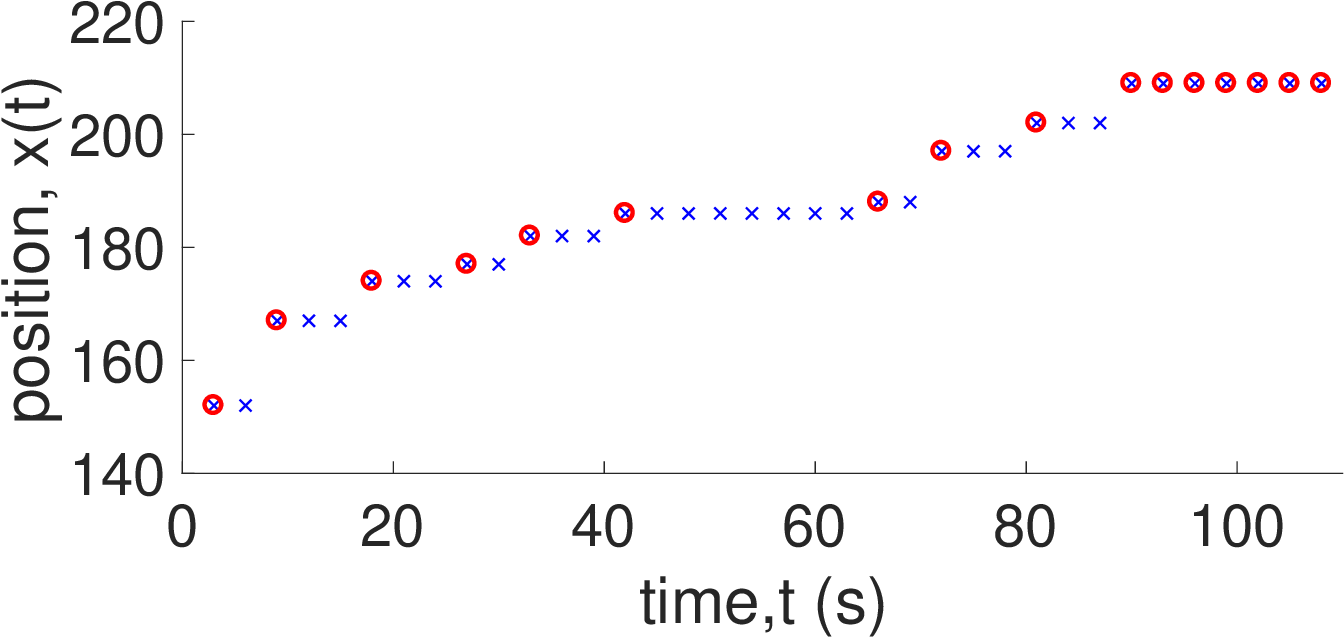}
         \caption{Removing repeated data points which are not at bus stops. Red circles indicates the remaining points.}
         \label{fig:inter2}
     \end{subfigure}
     \begin{subfigure}{.48\linewidth}
         \centering
         \includegraphics[width=\textwidth]{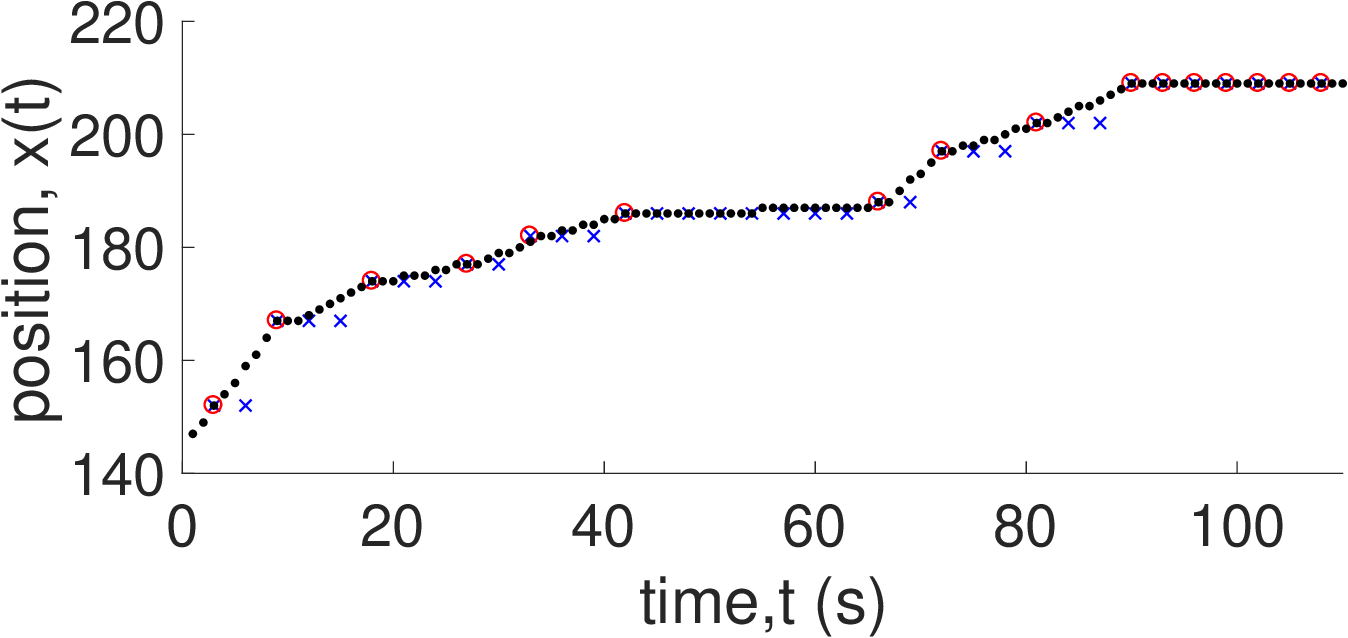}
         \caption{Black dots is the linear interpolation of the remaining points.}
         \label{fig:inter3}
     \end{subfigure}
     \begin{subfigure}{.48\linewidth}
         \centering
         \includegraphics[width=\textwidth]{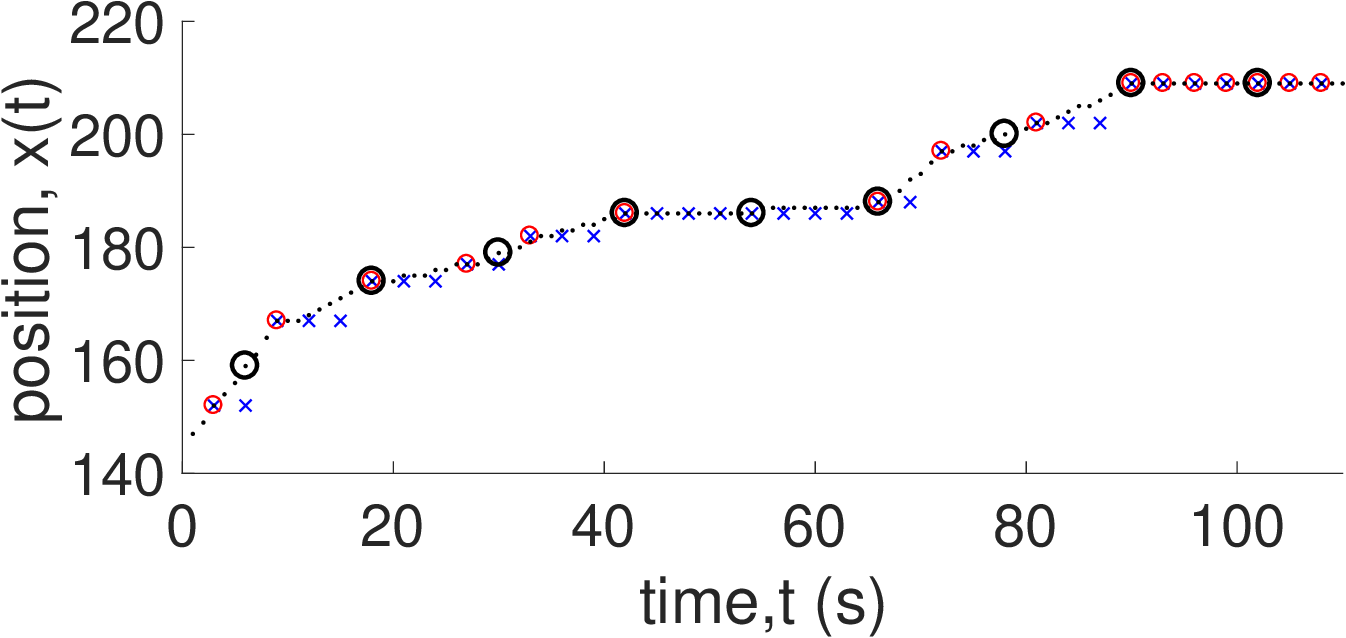}
         \caption{Black circles indicates the filtered data in a intended time resolution of $\delta t = 12s$}
         \label{fig:inter4}
     \end{subfigure}
        \caption{Illustration of the interpolation process done to filter out false occurrence of zero velocities.}
        \label{fig:interpolating}
\end{figure}

\begin{figure}[h!]
  \centering
  \includegraphics[width=\linewidth]{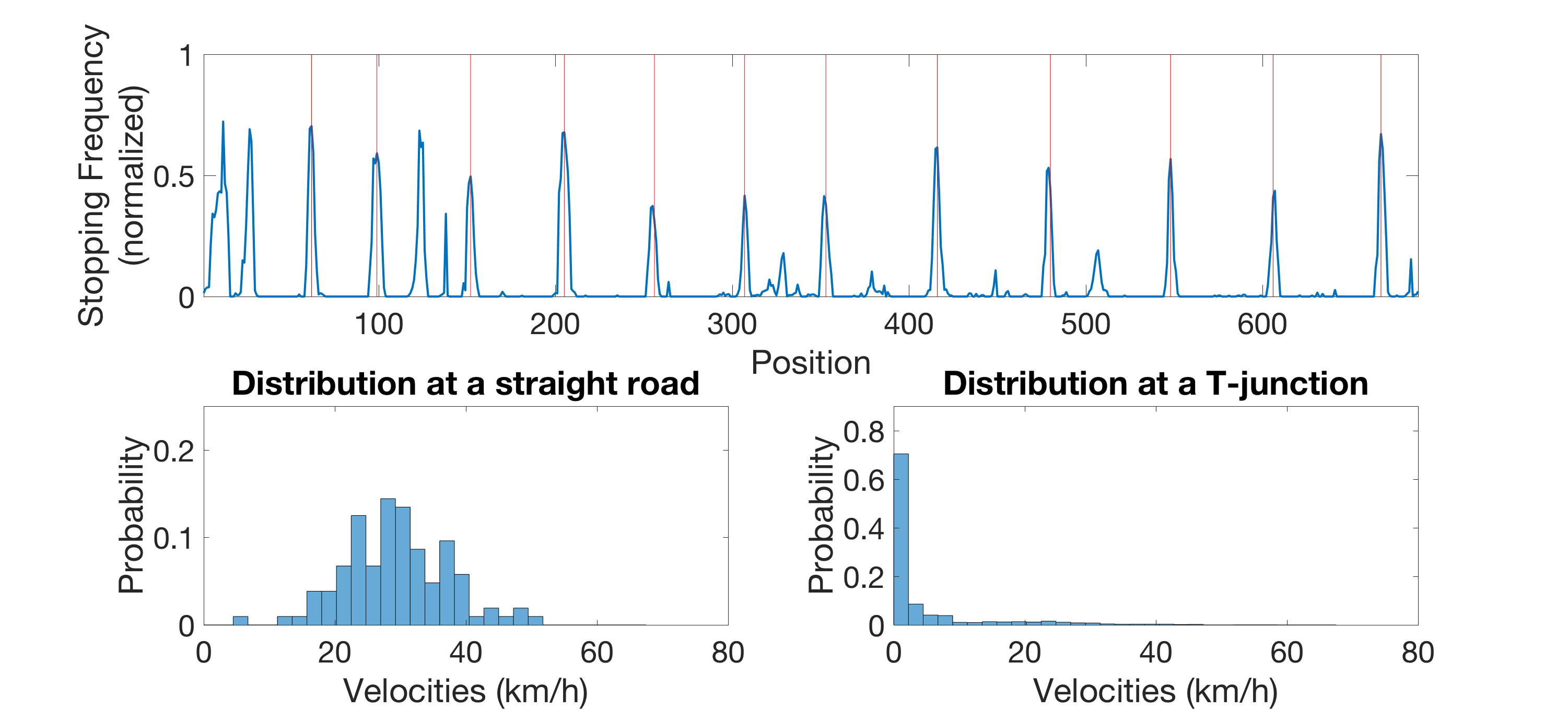}
  \caption{Graphs of normalized stopping frequency along the bus route, and velocity distributions of two different positions on the road. These graph are derived after the filtering process as described. Red lines on the graph are the sites where there are bus stops.}
  \label{fig:filteredvelocities}
\end{figure}

We assume that this artefact of zero velocities affect the road sites and not the bus stops. Figure \ref{fig:interpolating} illustrates the process of removing these false occurrences of zero velocities from the road sites. The data points in a single bus trajectory contains large amount of repeated positional data (Fig.\ref{fig:inter1}). This results from (1) buses being stationary, (2) buses experiencing interruption in updating their live positions, and (3) sampling the web portal at a higher frequency than its refresh rate. Regardless, all repeated positional data are removed (Fig. \ref{fig:inter2}), keeping only data which reflects when buses moved. The remaining positional data are linearly interpolated at a time interval of $\delta t = 1$s, based on the assumption that buses moves with a uniform speed between each positional update (Fig. \ref{fig:inter3}) \footnote{It was found that interpolating the data with a collection interval of $\delta t = 3$s leads to an unrealistic reduction of stoppages at the traffic junctions.}. These updated trajectories are then taken at the intended time resolution of $\delta t = 12$s (Fig. \ref{fig:inter4}). With the appropriate zero velocities removed, Fig. \ref{fig:filteredvelocities} illustrates the revised stopping frequency and velocity distribution of the same two sites as that of Fig. \ref{fig:unfilteredvelocities}.

Stoppages at bus stop sites are treated differently. As discussed, we assume that all stoppages at the bus stops are due to interactions with passengers, i.e, boarding and alighting, with a time delay $\tau$ used in calculating $\Delta t^{\text{arr}}_j$ (see Eq. (\ref{eqn:t_inj})). In the construction of the empirical velocity distribution at bus stops, only the non-zero velocities with which the bus departs the bus stop are considered. This ensures that buses move off when there are no more boarding or alighting of passengers.

%%% Insert biography
\bibliography{bib_mendeley,bib_others}
\bibliographystyle{unsrt}

\end{document}